\documentclass[twocolumn,trackchanges]{aastex7}
\usepackage{amsmath}
\usepackage{url}
\usepackage{hyperref}

\begin{document}
\title{The Study of Detailed Morphology of Milky Way Dwarf Galaxies: Draco and Bo\"otes I}
\author[orcid=0009-0002-9382-7068,sname='Sato']{Kyosuke S. Sato}
\affiliation{Astronomical Science Program, The Graduate University for Advanced Studies, SOKENDAI, 2-21-1 Osawa, Mitaka, Tokyo 181-8588, Japan}
\affiliation{National Astronomical Observatory of Japan, 2-21-1 Osawa, Mitaka, Tokyo 181-8588, Japan}
\email[show]{kyosuke.sato@grad.nao.ac.jp} 

\author[orcid=0000-0001-7550-2281,sname='Yagi']{Masafumi Yagi}
\affiliation{National Astronomical Observatory of Japan, 2-21-1 Osawa, Mitaka, Tokyo 181-8588, Japan}
\affiliation{Department of Advanced Sciences, Faculty of Science and Engineering, Hosei University, 3-7-2 Kajino-cho, Koganei, Tokyo 184-8584, Japan}
\email{masafumi.yagi@nao.ac.jp}

\author[orcid=0000-0002-7866-0514,sname='Okamoto']{Sakurako Okamoto}
\affiliation{National Astronomical Observatory of Japan, 2-21-1 Osawa, Mitaka, Tokyo 181-8588, Japan}
\affiliation{Astronomical Science Program, The Graduate University for Advanced Studies, SOKENDAI, 2-21-1 Osawa, Mitaka, Tokyo 181-8588, Japan}
\email{sakurako.okamoto@nao.ac.jp}

\begin{abstract}
We present a photometric study of the outer stellar structures of the Milky Way dwarf spheroidal galaxies (dSphs) Draco and Bo\"otes~I, using deep wide-field imaging data.
The old main-sequence stars are used to trace their spatial distributions beyond their tidal radii.
We derive global structural parameters and radial profiles of number density, ellipticity, and position angle from their stellar distributions.
Draco and Bo\"otes~I show contrasting features.
Draco shows no prominent extended stellar distribution, although weak elongations are detected toward the eastern and northwestern directions.
Its stellar distribution remains relatively compact compared with its estimated Jacobi tidal radius ($r_{\rm J}=57\arcmin.2^{+13.5}_{-11.1}$), suggesting that the stellar component is tightly bound by its gravitational potential.
In contrast, the radial number-density profiles of Bo\"otes~I along several directions show excess from the best-fit exponential profile outside of around $r\sim20'$, near the estimated Jacobi radius ($r_{\rm J}=21\arcmin.6^{+9.3}_{-7.6}$).
In addition, the ellipticity of Draco decreases with radius, whereas that of Bo\"otes~I increases toward larger radii.
Bo\"otes~I also exhibits a prominent S-shaped stellar structure extending approximately along its orbital direction.
Together, the radial density excess, increasing ellipticity, and S-shaped morphology suggest that the outer stellar structure of Bo\"otes~I may trace tidal tails produced by interactions with the Milky Way.
The contrasting morphologies of Draco and Bo\"otes~I further suggest that the response of dwarf galaxies to the Milky Way tidal field may depend not only on their orbital histories, but also on the internal structure of their dark-matter halos.
\end{abstract}

\section{Introduction}\label{sec:intro}
In the $\Lambda$CDM cosmological model, structure formation in the Universe proceeds hierarchically through the growth of initial dark matter density fluctuations (e.g., \citealp{1988ApJ...327..507F}). 
In this scenario, small systems form first and subsequently merge to build larger galaxies and galaxy clusters. 
As a result, the stellar halos of massive galaxies are expected to contain numerous substructures that trace past accretion and merging events (e.g., \citealp{1978ApJ...225..357S,2001ApJ...558..666B,2005ApJ...635..931B,2010MNRAS.406..744C}).
Indeed, many such remnants have been discovered in the Milky Way (MW) halo as stellar streams and kinematic substructures (e.g., \citealp{2018Natur.563...85H,2021ApJ...920...51M,2022Natur.601...45M,2023A&A...673A..44F}).

While hierarchical assembly scenario is well established on the scale of the MW halo, it remains unclear to what extent similar processes operate at much smaller mass scales. 
\citet{2014ApJ...794..115D} pointed out that approximately $45$--$70\%$ of surviving dwarf galaxies experienced at least one major dwarf--dwarf merger over their histories, where major mergers were defined by a stellar mass ratio of $M_{\star,2}/M_{\star,1}\gtrsim0.1$ (see Figure 1 of \citealp{2014ApJ...794..115D}).
Minor mergers below this threshold were not included in their merger fractions.
Most of these major mergers occurred at early times.
Since $z\sim1$, the merger fraction decreases to about $10\%$ for satellite dwarfs with $M_\star>10^6\,M_\odot$ and to approximately $5\%$ or less for lower-mass systems.
They also found that the merger fraction decreases toward lower stellar masses, suggesting that dwarf-dwarf mergers become progressively rarer at the lowest galaxy mass scales.

Several studies have reported extended stellar distributions and substructures around Local Group dwarf galaxies, for which both merger-driven and tidal origins have been proposed \citep[e.g.,][]{2021NatAs...5..392C,2021ApJ...923..218F,2023MNRAS.525.2875S,2024MNRAS.527.4209J,2025ApJ...993L...7S}.
\citet{2024MNRAS.527.4209J} searched for extended stellar distributions around $\sim60$ MW dwarf satellites and identified nine systems showing evidence for an extended outer stellar component.
This relatively high occurrence of extended structures ($\sim15\%$) persists down to the ultra-faint dwarf regime, which may be difficult to be produced by the dwarf-dwarf mergers only.
The extended stellar structures can also be produced by tidal interactions with the MW \citep{2009ApJ...698..222P}.
Searching for and characterizing extended structures in low-mass dwarf galaxies is important for observationally constraining the role of mergers at the lowest galaxy mass scales.

In this context, \citet{2025ApJ...993L...7S} reported that the extended stellar distribution of the Ursa Minor dwarf spheroidal galaxy (dSph) is oriented approximately perpendicular to its orbital direction.
This morphology is difficult to explain solely as a tidal feature aligned with the orbit and raises the possibility of a merger-related origin even at the relatively low stellar mass of Ursa Minor ($M_\star \simeq 2.9\times10^{5}\,M_\odot$; \citealp{2012AJ....144....4M}).
The work demonstrated that detailed measurements of morphology and stellar dynamics of low-mass galaxies are essential for determining their formation origins.

In this study we investigated two Galactic dSphs, Draco and Bo\"otes I.
Draco is one of the classical dSph galaxies in the MW satellites, with a stellar mass of $2.9\times10^5\,M_\odot$ \citep{2012AJ....144....4M}. 
This value places Draco close to the boundary between classical dSph galaxies and ultra-faint dwarf galaxies (UFDs). 
It is a metal-poor system with a mean stellar metallicity of $\langle{\rm [Fe/H]}\rangle = -1.93 \pm 0.01$ dex \citep{2011ApJ...727...78K}. 
Draco is located at a heliocentric distance of $\sim75.8\pm0.7({\rm stat})\pm5.4({\rm sys})\ {\rm kpc}$ \citep{2004AJ....127..861B}. 
Dynamical studies have shown that Draco is strongly dominated by dark matter \citep{2002MNRAS.330..792K,2024ApJ...970....1V}.
While the presence of binary stars can potentially introduce systematic biases into velocity dispersion measurements in stellar systems \citep[e.g.,][]{2022ApJ...939....3P}, recent high-precision dynamical modeling by \citet{2024ApJ...970....1V} demonstrates that the high binary fraction does not significantly alter the inferred velocity dispersion or the high mass-to-light ratio of Draco.
Because of its large velocity dispersion and apparently undisturbed morphology, Draco has long been considered one of the most dynamically stable dSph galaxies in the MW halo.

However, based on the spectroscopic data of the Dark Energy Spectroscopic Instrument (DESI), \citet{2025ApJ...994..134D} recently reported that eight high-probability Draco member stars are located beyond the nominal tidal radius.
Their result suggests that Draco has experienced a moderate level of tidal influence from the MW, intermediate between Sculptor and Fornax. 
However, the spatial distribution of the outer stars is not aligned with the orbital direction expected for tidally stripped material.
They therefore suggested that these stars may be associated with an extended stellar component, potentially originating from past dwarf--dwarf merger events.

Bo\"otes~I is one of the ultra-faint dwarf (UFD) galaxies around the MW, discovered in the Sloan Digital Sky Survey \citep{2006ApJ...647L.111B}. 
It has a stellar mass of $M_\star \sim 2.9\times10^4\,M_\odot$ and is located at a heliocentric distance of $66\pm3$ kpc \citep{2006ApJ...653L.109D, 2012AJ....144....4M}.
Bo\"otes~I hosts a metal-poor stellar population with a mean metallicity of $\langle{\rm [Fe/H]}\rangle \sim -2.53\pm0.05$ \citep{2025arXiv251201547M}. 
Recent photometric studies have revealed evidence for extended stellar structures around Bo\"otes~I. 
For example, \citet{2021ApJ...923..218F} examined the spatial distribution of blue horizontal branch (BHB) and blue straggler stars (BSS) and found that candidate members extend beyond 10 times of half-light radius ($r_{\rm h}$), suggesting that Bo\"otes~I may possess an extended stellar component.
More recent analyses combining photometric and spectroscopic data have suggested that the outer stellar distribution of Bo\"otes~I may contain tidal features aligned with its orbital motion around the MW \citep{2022MNRAS.516.2348L,2026ApJ...998...47S}. 
These results suggest that the outer structure of Bo\"otes~I may preserve signatures of its dynamical evolution within the MW halo.
While tidal interactions with the MW provide a natural explanation for extended stellar structures, alternative scenarios have also been proposed.
\citet{2023MNRAS.519.1349W} found that several outer metal-poor stars are consistent with being tidally stripped from the main body, but also noted that their chemical similarity could indicate formation in a separate low-mass dwarf galaxy that later merged with Bo\"otes~I.
Therefore, the origin of the outer stellar population remains under debate, with both tidal stripping and a past dwarf--dwarf merger providing plausible explanations.
Constraining between these scenarios requires detailed measurements of the morphology and stellar populations in the outskirts of Bo\"otes~I.

In this work, we therefore use deep and wide-field photometric data to investigate the spatial distribution of Draco and Bo\"otes~I.
This paper is organized as follows. In section \ref{sec:data}, the details of data reduction are summarized. 
In section \ref{sec:property}, the structural analysis of the
Draco and Bo\"otes I are described. 
In section \ref{sec:radial_profile}, we describe the analytical results of radial profiles for Draco and Bo\"otes I. 
Section~\ref{sec:discussion} discusses the implications of our results.
We also estimate the Jacobi tidal radius of Ursa Minor, previously studied by \citet{2025ApJ...993L...7S}, and compare its outer stellar structure with those of Draco and Bo\"otes~I.
Conclusion is given in Section \ref{sec:concl}
\setcounter{footnote}{0}
\section{Data and reduction}\label{sec:data}
\subsection{Data}
    \begin{figure*}[ht!]
        \begin{center}
        \includegraphics[width=16cm]{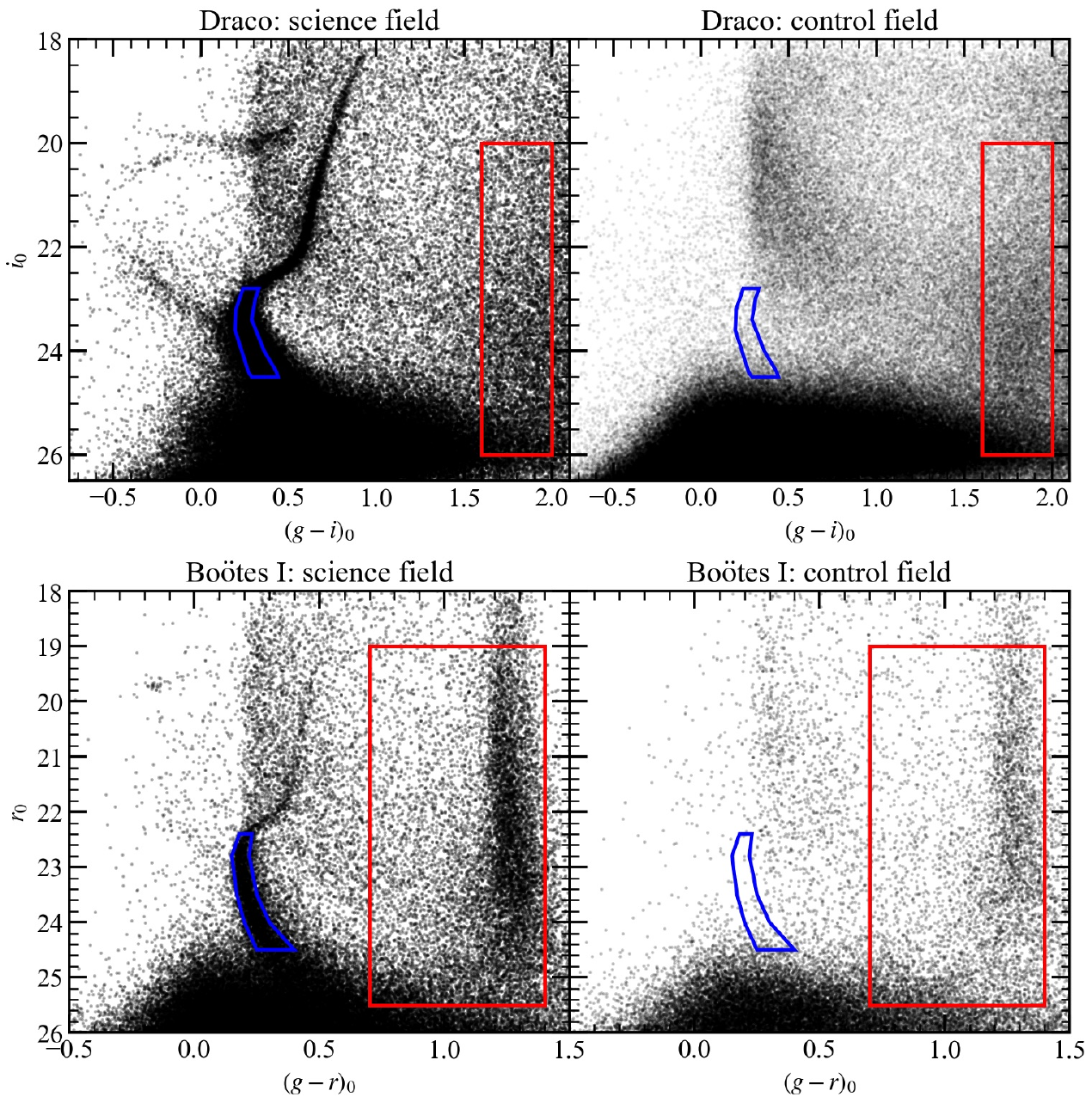}
        \end{center}
        \caption
            {
            The CMDs of Draco and Bo\"otes I are shown in the left panels, and those of the corresponding control fields are shown in the right panels. 
            Because the science and control fields cover different areas, the control-field sources are plotted with reduced opacity; the alpha values are scaled by factors of $1/4$ for Draco and $1/1.4$ for Bo\"otes I. 
            The blue polygons define the CMD selection regions for candidate MS member stars. 
            The red rectangles define foreground-dominated regions used to estimate the magnitude-dependent detection completeness of the control-field catalogs. 
            This completeness is inferred by comparing the number counts of point sources in the control fields with the completeness-corrected number count of the corresponding sources in the science fields.
            }\label{fig:CMD}
    \end{figure*}
Photometric observations of the Draco and the Bo\"otes I dwarf spheroidal galaxies were conducted using the Hyper Suprime-Cam (HSC; \citealp{2018PASJ...70S...1M, 2018PASJ...70S...2K, 2018PASJ...70...66K, 2018PASJ...70S...3F}) on the Subaru Telescope. 
The data used in this study are retrieved from the Subaru-Mitaka-Okayama-Kiso-Archive (SMOKA).
Draco was observed on 2016 April 3 (UTC) in the $g$- and $i_2$-bands\footnote{\label{fn:i2}The $i$-band filter of Subaru/HSC was replaced with the $i_2$ filter in February 2016. 
Throughout this paper, we denote the $i_2$-band simply as $i$.} (proposal ID: S16A-TE134), 
while Bo\"otes~I was observed on 2014 July 2, 3, and 2015 May 24 in the $g$- and $r$-bands (proposal ID: S14A-191, S15A-092).
The details of observing conditions are summarized in Table \ref{tab:parameters}.
These excellent photometric conditions and sufficient exposure times allow us to reach the old main-sequence turn-off (MSTO) and to obtain numerous main-sequence (MS) member stars.
Throughout this paper, all equatorial coordinates are given for equinox J2000.0.
\begin{table*}[ht!]
    \caption{Information about the Data}\label{tab:parameters}
    \centering
    \footnotesize
    \renewcommand{\arraystretch}{1.5}
    \begin{tabular*}{12cm}{@{\extracolsep{\fill}}cccccc}
        \hline
        field & Filter & R.A. (J2000) & Decl. (J2000) & Seeing & Exposure time \\
        \hline
        Draco\_F1 &
        $g$ &
        $17^\mathrm{h}17^\mathrm{m}53\fs02$ &
        $\mathrm{+57^{\circ}52\arcmin14\farcs15}$ &
        $0\farcs41$--$0\farcs45$ &
        $200\,\mathrm{s}\times6$ \\

        Draco\_F2 &
        $g$ &
        $17^\mathrm{h}22^\mathrm{m}31\fs91$ &
        $\mathrm{+57^{\circ}57\arcmin26\farcs43}$ &
        $0\farcs46$--$0\farcs58$ &
        $200\,\mathrm{s}\times6$ \\

        Draco\_F1 &
        $i2$ &
        $17^\mathrm{h}17^\mathrm{m}53\fs02$ &
        $\mathrm{+57^{\circ}52\arcmin14\farcs16}$ &
        $0\farcs55$--$0\farcs67$ &
        $200\,\mathrm{s}\times9$ \\

        Draco\_F2 &
        $i2$ &
        $17^\mathrm{h}22^\mathrm{m}31\fs93$ &
        $\mathrm{+57^{\circ}57\arcmin26\farcs44}$ &
        $0\farcs58$--$0\farcs65$ &
        $200\,\mathrm{s}\times9$ \\
        \hline

        Bo\"{o}tes I &
        $g$ &
        $14^\mathrm{h}00^\mathrm{m}04\fs86$ &
        $\mathrm{+14^{\circ}30\arcmin10\farcs84}$ &
        $0\farcs48$--$1\farcs03$ &
        $120\,\mathrm{s}\times39$ \\

        Bo\"{o}tes I &
        $r$ &
        $14^\mathrm{h}00^\mathrm{m}04\fs87$ &
        $\mathrm{+14^{\circ}30\arcmin10\farcs86}$ &
        $0\farcs53$--$1\farcs00$ &
        $120\,\mathrm{s}\times13$ \\
        \hline
    \end{tabular*}
\end{table*}
\subsection{Reduction and Photometry}\label{reduction}
The data are processed with version 8.5.3 of the HSC data reduction pipeline ($\rm{hscPipe}$; \citealt{2018PASJ...70S...5B}). 
The pipeline performs standard CCD-level processing, including bias and dark subtraction, flat-field correction, and sky subtraction, on individual CCD frames. 
Astrometric solutions and photometric calibration are then obtained by matching bright sources in each frame to the Pan-STARRS1 catalog \citep{2012ApJ...750...99T, 2016arXiv161205560C}, following the procedure described in \cite{2018PASJ...70S...6H}. 
After calibration, the individual exposures are stacked based on mosaic images for each band, during which cosmic rays are identified and removed. 

Photometric measurements are performed on the co-added images using both point-spread-function (PSF) and CModel photometry \citep{2018PASJ...70S...5B}. 
In $\rm{hscPipe}$, the PSF is estimated with PSFEx \citep{2011ASPC..442..435B, 2013ascl.soft01001B}. 
To  resolved sources from extended objects, we use the \texttt{Extendedness} parameter estimated by the pipeline, which quantifies the difference between PSF and CModel magnitudes \citep{2018PASJ...70S...5B}. 
This selection is used to define the stellar sample adopted throughout our analysis.

Corrections for the Galactic extinction are applied based on the dust reddening maps of \citet{2011ApJ...737..103S}. 
The $E(B-V)$ values across the observed fields are obtained from the NASA/IPAC Infrared Science Archive (IRSA)\footnote{https://irsa.ipac.caltech.edu/applications/DUST/}. 
To derive the band-dependent extinction coefficients for the Subaru/HSC $g$-, $r$-, and $i$-band filter systems, we perform synthetic photometry using a stellar atmosphere model from Castelli \& Kurucz \citep{2003IAUS..210P.A20C}.
We adopt a model with an effective temperature of $T_{\rm eff}=5500,\mathrm{K}$ and a surface gravity of $\log g = 4.0$.
To account for the wavelength-dependent response of Subaru/HSC, synthetic fluxes are calculated by integrating the product of the model spectrum and the total system throughput. 
The system throughput includes the atmospheric transmission, the transmission and reflectivity of the telescope and instrument optics, the filter transmission, and the CCD quantum efficiency\footnote{\url{https://www.subarutelescope.org/Observing/Instruments/HSC/index.html}} \citep{2018PASJ...70...66K, 2018PASJ...70S...1M}.
We then apply the Galactic extinction curve of \citet{1999PASP..111...63F} to the model spectrum assuming $E(B-V)=1$ and calculate synthetic magnitudes for both the intrinsic and reddened spectra in each HSC band.
The band-dependent extinction coefficients are obtained directly from the difference between the reddened ($m_{X,\rm red}$) and intrinsic synthetic magnitudes ($m_{X,\rm int}$),
\begin{equation}
\frac{A_X}{E(B-V)}
=
m_{X,\rm red}
-
m_{X,\rm int},
\qquad
X \in \{g,r,i\},
\end{equation}
where the equality follows from the adopted $E(B-V)=1$.
The extinction coefficients are derived as,
    \begin{align}
    g_{0} &= g - 3.111E(B-V), \\
    r_{0} &= r - 2.238E(B-V),\ \text{and}\\
    i_{0} &= i - 1.603E(B-V).
    \label{extinction}
    \end{align}
Here, $g_0$, $r_0$, and $i_0$\textsuperscript{\ref{fn:i2}} denote the extinction-corrected magnitudes in the $g$-, $r$-, and $i$-bands, respectively. 
The extinction correction is applied to each star individually using the $E(B-V)$ value at its celestial position.
The mean Galactic extinction values are 0.05 and 0.04 mag in the $g$- and $r$-bands for Bo\"otes I, and 0.09 and 0.05 mag in the $g$- and $i$- bands for Draco, respectively.
The color--magnitude diagram (CMD) of point source in Draco and Bo\"otes I are shown in the left panels of Figure \ref{fig:CMD}.
\subsection{Detection Completeness}
To quantify the detection completeness of point sources in our photometric catalog, we perform artificial star injection tests following the procedure described in \citet{2025MNRAS.536..530O}. 
For this artificial test, we follow the same procedure of \cite{2025PASJ...77.1259S}, and estimate the parameters of $m_{50}$, $A$, and $\rho$ by fitting to the completeness function $\eta(m)$, described as
\begin{equation}
    \eta(m)
    =
    \frac{A}
    {1+\exp(\frac{m-m_{50}}{\rho})}.
    \label{eq:completeness}
\end{equation}
$m_{50}$ is the magnitude at which the completeness reaches 50\%, and $A$ and $\rho$ are the normalization and the parameter of the slope, respectively.
For Draco, the field is partitioned into a $7\times7$ grid, while for Bo\"otes I a $6\times6$ grid is adopted, reflecting differences in the spatial coverage of the HSC observations. 

In the case of the Draco, the detection completeness generally decreases toward the central regions of the galaxy, due to increased stellar crowding. 
The mean and root mean squares of 50\% completeness magnitudes over the entire field are $m_{g,50}=25.68 \pm 0.51$ mag and $m_{i,50}=24.67 \pm 0.37$ mag.
For Bo\"otes I, the detection completeness exhibits a smaller spatial variation across the field compare to Draco.
The corresponding mean 50\% completeness magnitudes are $m_{g,50}= 25.05\pm0.25$ mag and $m_{r,50}= 24.79\pm0.36$ mag.
The completeness as a function of color and magnitude is then evaluated locally in each subregion, and the resulting values are used to assign weights to individual stars in the CMDs.

Since we have no data next to the science fields, we estimate the contamination using control fields selected from the Hyper Suprime-Cam Subaru Strategic Program Public Data Release 3 (PDR3) Wide \citep{2022PASJ...74..247A}. 
These control fields have photometric depths comparable to those of the science fields and are located at Galactic latitudes similar to those of the target dwarf galaxies.
The contamination is estimated from rectangular control fields defined as $140^{\circ} \leq \mathrm{R.A.} \leq 144^{\circ}$ and $0^{\circ} \leq \mathrm{Dec.} \leq +3^{\circ}$ for Draco, and $210^{\circ} \leq \mathrm{R.A.} \leq 211^{\circ}$ and $+42.5^{\circ} \leq \mathrm{Dec.} \leq +44.5^{\circ}$ for Bo\"otes I. 
Both Draco and Bo\"otes~I are located at relatively high Galactic latitudes, with $b=+34.72^\circ$ and $b=+69.23^\circ$, respectively. 
As a result, contamination from the MW disk is expected to be small in our analysis (see the sources within the blue polygon in the right panels of Figure \ref{fig:CMD}).

To account for the detection completeness of control fields, we derive a magnitude-dependent correction using sources in the color-magnitude region where the dSph’s member stars are assumed to be absent (the red boxes in Figure \ref{fig:CMD}).
For each $g-$ and $i-$bands, we construct the magnitude distribution of the selected foreground sources in the science field after applying the detection-completeness correction, and compare it with the unweighted number count of the corresponding sources in the control field. 
We then fit the ratio of the unweighted control-field number count to the completeness-corrected science-field number count using the functional form in Equation (2) of \cite{2025PASJ...77.1259S}, independently for the $g-$, $r-$ and $i-$bands. 
\begin{table}
\centering
\caption{Best-fitting parameters for the magnitude-dependent completeness correction of the control fields.}
\label{tab:control_completeness}
\setlength{\tabcolsep}{10pt}
\begin{tabular}{lcccc}
\hline
Field & Band & $A$ & $\rho$ & $m_{50}$ \\
\hline
Draco\_CF      & $g$ & 2.143 & 0.621 & 25.345 \\
Draco\_CF      & $i$ & 0.915 & 0.670 & 24.207 \\
Bo\"otes~I\_CF & $g$ & 0.478 & 1.707 & 24.034 \\
Bo\"otes~I\_CF & $r$ & 0.910 & 0.953 & 24.767 \\
\hline
\end{tabular}
\end{table}
The best-fitting parameters for the magnitude-dependent completeness correction of the control fields are summarized in Table~\ref{tab:control_completeness}.
Using these parameters, we estimate the weights of each object of control fields.

\section{Property of Draco and Bo\"otes I}\label{sec:property}
\subsection{Main Sequence Stars Selection}\label{sec:MSselection}
We select MS stars down to $i_0=24.5$ for Draco and $r_0=24.5$ for Bo\"otes~I using the blue polygons shown in Figure~\ref{fig:CMD}.
The MS-selection polygons are defined to maximize the fraction of member stars while minimizing contamination from MW foreground stars and background galaxies.
Their shapes and magnitude ranges are determined based on the CMD distributions of both the science and control fields, with particular emphasis on obtaining a clean sample in the low-density outer regions.
Although the full-field CMD top-left panel of Figure \ref{fig:CMD} of Draco shows a relatively broad color distribution, the MS distribution becomes narrower at larger radii, as shown in Figure~\ref{fig:CMD_tile}.
We therefore adopt a relatively narrow selection polygon that follows the outer-region MS sequence, which helps reduce foreground and background contamination in the analysis of the outer stellar structure.

Since the MS populations of these dSphs are generally old and metal-poor, they occupy relatively blue regions of the CMD compared with most MW foreground stars.
Consequently, foreground contamination is less severe in the adopted MS-selection regions than in the RGB selection regions used in many previous studies, where proper-motion information is often required for reliable member selection.
    \begin{figure*}[ht!]
        \begin{center}
        \includegraphics[width=14cm]{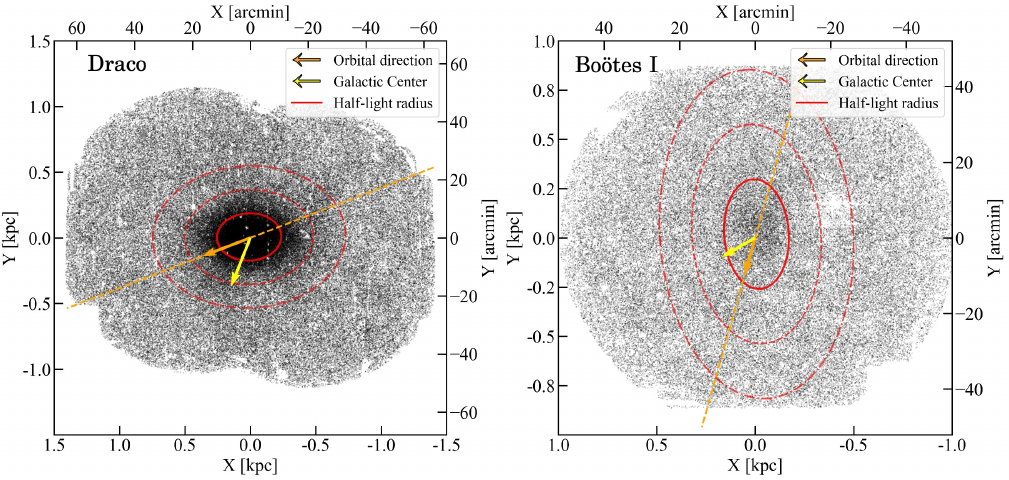}
        \end{center}
        \caption
        {
        Spatial distributions of all point-sources of Draco dSph (top-left) and Bo\"otes I (bottom-right).
        In the spatial distribution, the orbital directions are indicated as the orange arrow and dashed line, direction to the MW center as yellow arrow.
        The 1, 2, and 3 times half-light radius are shown as red solid, dashed, and dash-dotted ellipse, respectively.
        }\label{fig:spatial_dist}
    \end{figure*}
As shown in the right panels of Figure~\ref{fig:CMD}, only a small number of contaminants overlap with the adopted MS-selection regions
in the control fields.
In the following subsections, we use the stars selected within these regions in the left panels as member candidates to derive the structural properties of each dSph.

\subsection{Structural-parameter estimation}\label{sec:structure_param}
We estimate the structural parameters of Draco and Bo\"otes~I using the Bayesian MCMC method described in \citet{2025PASJ...77.1259S}.
Briefly, we fit a two-dimensional exponential model to the projected spatial distribution of the member candidates selected using the blue polygon in Figure~\ref{fig:CMD}.
When converting equatorial coordinates to projected coordinates, we adopt the literature centers $(\mathrm{R.A.}_0,\mathrm{Decl.}_0) =(17^\mathrm{h}20^\mathrm{m}12\fs4, +57^{\circ}54\arcmin55\farcs0)$ for Draco \citep{2014MNRAS.443.1151N} and $(14^\mathrm{h}00^\mathrm{m}06\fs0, +14^{\circ}30\arcmin00\farcs0)$ for Bo\"otes~I \citep{2006ApJ...647L.111B}.
We adopt distance moduli of $(m-M)_0=19.40$ mag for Draco \citep{2004AJ....127..861B} and $(m-M)_0=19.1$ mag for Bo\"otes~I \citep{2006ApJ...653L.109D}.
Here, the subscript 0 for distance moduli denotes the extinction-corrected.
These values are used to convert a separation angle into a physical distance.
    \begin{table*}[ht!]
    \caption{The structural parameters of the Draco and Bo\"otes~I dSphs}
    \label{table:post}
    \centering
    \footnotesize
    \renewcommand{\arraystretch}{1.5}

        \begin{tabular*}{\textwidth}{@{\extracolsep{\fill}}llllllll}
            \hline

            Galaxy &
            $\alpha_{0}\,\rm(J2000)$ &
            $\delta_{0}\,\rm(J2000)$ &
            $r_{\rm h}\,[\rm pc]$ &
            $r_{\rm h}\,[\arcmin]$ &
            $\epsilon$ &
            $\theta\,[^{\circ}]$ &
            $N_{\ast}$ \\

            \hline

            Draco &
            $17^{\rm h}20^{\rm m}15.58^{\rm s}\,
            {}^{+1.18^{\rm s}}_{-1.13^{\rm s}}$ &
            $+57^{\circ}55^{\prime}10.56^{\prime\prime}\,
            {}^{+3.56^{\prime\prime}}_{-3.46^{\prime\prime}}$ &
            $239.59^{+1.97}_{-1.90}$ &
            $10.84\pm0.09$ &
            $0.266\pm0.007$ &
            $92.37\pm0.89$ &
            $14673.89^{+14.24}_{-15.91}$ \\

            Bo\"otes~I &
            $14^{\rm h}00^{\rm m}04.90^{\rm s}\,
            {}^{+0.62^{\rm s}}_{-0.64^{\rm s}}$ &
            $+14^{\circ}30^{\prime}59.76^{\prime\prime}\,
            {}^{+15.48^{\prime\prime}}_{-15.84^{\prime\prime}}$ &
            $282.79^{+5.84}_{-6.02}$ &
            $14.73^{+0.30}_{-0.31}$ &
            $0.428^{+0.014}_{-0.015}$ &
            $3.12^{+1.30}_{-1.24}$ &
            $2652.21^{+36.98}_{-37.24}$ \\

            \hline
        \end{tabular*}
    \end{table*}
The fitting procedure is identical to that of \citet{2025PASJ...77.1259S}, except for the adopted position-angle ranges.
We use $0^\circ\leq\theta<180^\circ$ for Draco and $-90^\circ\leq\theta<90^\circ$ for Bo\"otes~I.
The latter range is adopted to prevent the posterior distribution from lying across the boundary of the fitting interval.
The completeness weights derived from the artificial-star tests are not applied in this structural fit.
However, we confirmed that the structural parameters obtained with and without the completeness weights are consistent within $1\sigma$ for both Draco and Bo\"otes~I.

The fitted parameters are the center coordinates $(\alpha_0,\delta_0)$, half-light radius $r_{\rm h}$, ellipticity $\epsilon$, position angle $\theta$, and number of selected member stars $N_\ast$. 
The resulting parameters are summarized in Table~\ref{table:post}.
Using the estimated parameters, we plot the half-light radii as red solid ellipses over the spatial distribution of Draco and Bo\"otes I dSphs in Figure \ref{fig:spatial_dist}. 

    \begin{figure*}[ht!]
        \begin{center}
        \includegraphics[width=17cm]{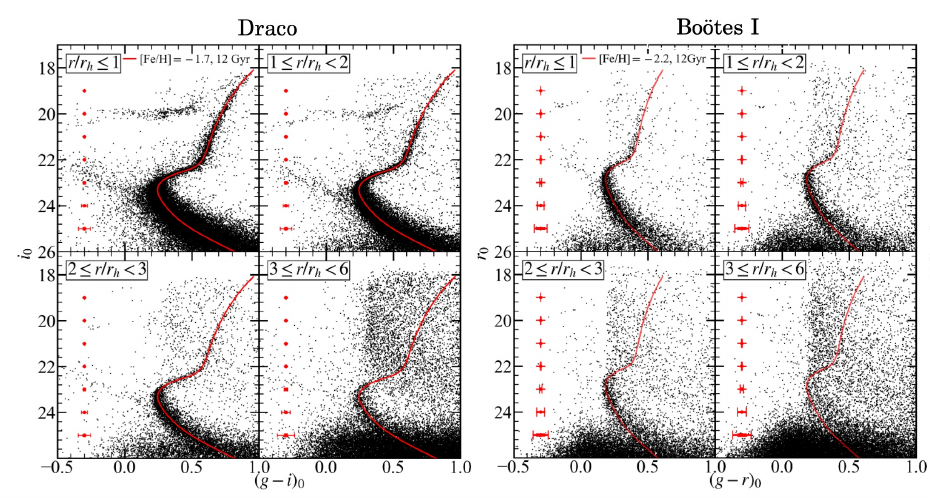}
        \end{center}
        \caption
        {
        The CMDs according to the distance from the center of each dSph. 
        Left panels are the CMDs of Draco and the right panels are of Bo\"otes I dSph.
        The red lines in CMDs are the BaSTI isochrones, which are shifted to the distance of each dSph (\citealp{2021ApJ...908..102P}; $\rm{[Fe/H]}=-1.7$ and 12 Gyr for Draco, and $\rm{[Fe/H]}=-2.5$ and 12.5 Gyr for Bo\"otes I).
        The red error bars show the photometric errors for magnitudes and colors.
        }\label{fig:CMD_tile}
    \end{figure*}

\subsection{Morphology of Draco dSph}\label{sec:morphologyDraco}
Previously, no prominent tidal features have been reported around Draco in previous photometric surveys \citep{2001AJ....122.2538O, 2007MNRAS.375..831S, 2018ApJ...860...66M}.
On the other hand, \citet{2005ApJ...631L.137M} show a flat velocity dispersion profile of Draco even beyond the nominal tidal radius ($40\arcmin.1$; \citealp{2001AJ....122.2538O}).
This flat slope has been interpreted as either evidence for a dark matter halo more extended than the stellar component, or as a consequence of tidal heating that inflates the outer velocity dispersion.

The Draco dSph is considered to have a more circular shape than other dSphs (e.g., $\epsilon = 0.30 \pm 0.01$; \citealp{2018ApJ...860...66M}).
Our estimated ellipticity, $\epsilon = 0.266 \pm 0.007$, is slightly more circular than the value reported by \citet{2018ApJ...860...66M}.
Using a broader main-sequence selection polygon that includes potentially younger and/or more metal-rich populations gives a slightly smaller half-light radius, differing by about $1.5\sigma$ from our fiducial value, while the ellipticity remains consistent within $1\sigma$.
Our data covers beyond the nominal tidal radius ($40\arcmin.1$; \citealp{2001AJ....122.2538O}) and also deep enough to detect intrinsically faint MS stars.
This large number of MS population allows us to investigate the detailed outskirts structure.

\begin{figure*}[ht!]
        \begin{center}
        \includegraphics[width=16cm]{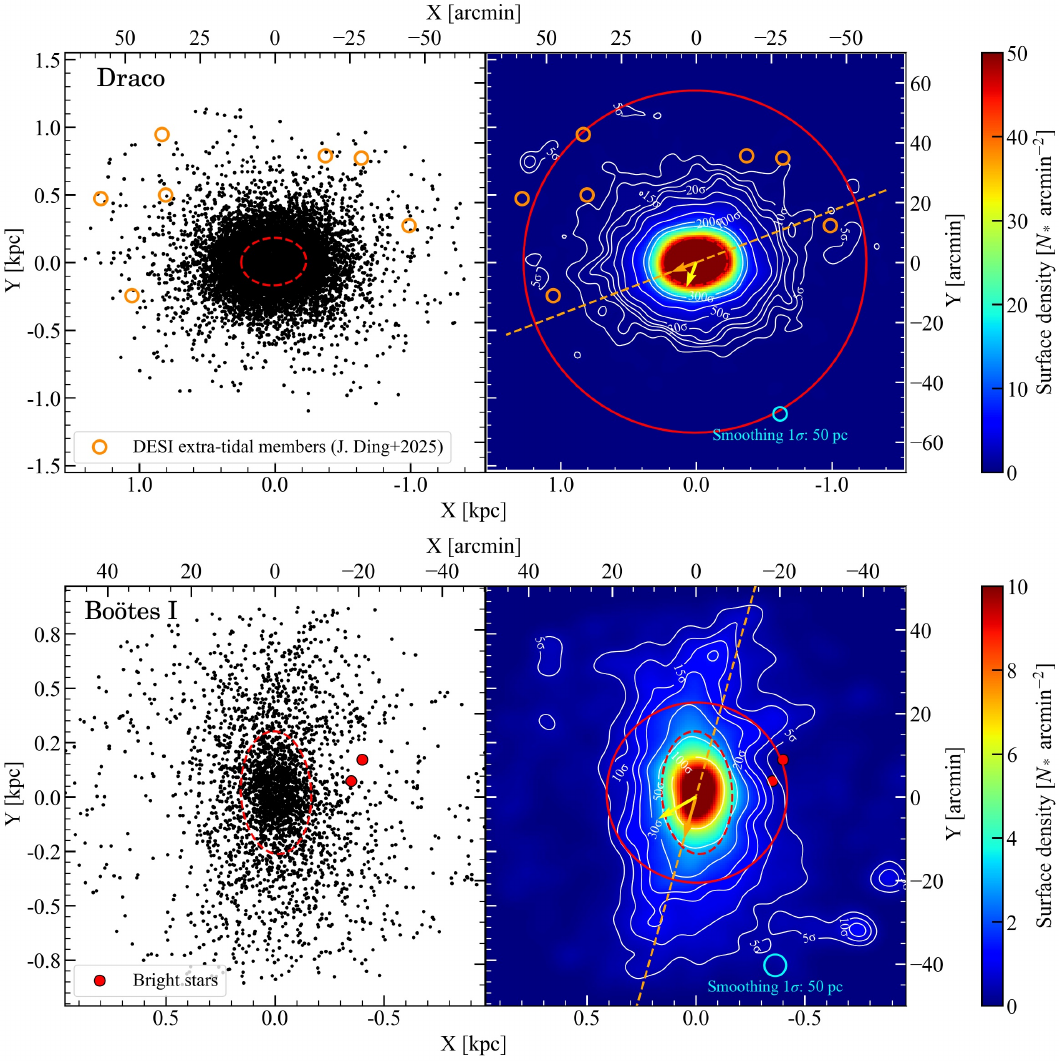}
        \end{center}
        \caption
        {
        Left panels display the spatial distributions of the selected MS stars (see Figure~\ref{fig:CMD}), and the right panels show the stellar surface density maps of the Draco (top) and Bo\"{o}tes I (bottom) dSphs.
        We set the $200\times 200$ bins for surface density map.
        The contours indicate the isodensity levels measured in the $\sigma$ from the background, with the 5, 10, 20, 30, 50, 100, 200, 300 $\sigma$ contour for Draco and 5, 10, 20, 30, 50, 100 $\sigma$ contour for Bo\"otes I.
        The red dashed ellipses in both the left and right panels indicate the half-light radius.
        The red solid circles indicate the Jacobi tidal radius.
        The dashed orange lines and orange arrows indicate the orbital direction as same as Figure \ref{fig:spatial_dist}.
        The orange small circles are the positions of extra tidal stars discovered by \cite{2025ApJ...994..134D}.
        The two red dots indicates foreground bright stars $G_{\rm{gaia}}\lesssim7$ magnitude, which interfere with the detection of faint stars.
        }\label{fig:contour}
    \end{figure*}

To compare the radial differences of stellar populations in Draco, we divided the observed footprint into four regions using the derived structural parameters.  
The CMDs of point sources within each region are shown in the left panels of Figure \ref{fig:CMD_tile}.
For comparison, we overlay a BaSTI isochrone (\citealp{2021ApJ...908..102P}; $[\mathrm{Fe/H}]=-1.7$, $[\alpha/\mathrm{Fe}]=0.4$, age $=12$~Gyr) shifted to the distance of Draco.
Figure~\ref{fig:CMD_tile} shows that the innermost region ($r/r_{\rm h} \leq 1$) exhibits a broader and bluer MSTO than the outer regions.
This difference may reflect variations in stellar age, metallicity, and/or $[\alpha/\mathrm{Fe}]$, since these parameters can all affect the position and width of the MSTO.
A detailed analysis of the differences in the star formation histories among these regions will be presented elsewhere.
We show the spatial distribution and Gaussian-smoothed surface-density maps of MS stars of Draco in the upper panels of Figure \ref{fig:contour}. 
A Gaussian kernel with $\sigma=50$ pc is used to construct the surface-density maps. 
The background mean density $(\mu_{\rm bg})$ and its standard deviation $(\sigma_{\rm bg})$ are estimated from the lower half of the pixel-value distribution in the smoothed density map. 
The contours are then drawn at levels of $\mu_{\rm bg}+n\sigma_{\rm bg}$.

To assess the possible influence of the MW tidal field on the morphology of each dSph, we estimate the Jacobi tidal radius \citep{2008gady.book.....B}. 
We adopt the simplified two-body approximation,
\begin{equation}
    \label{eq:jacobi_radius}
    r_{\rm J}
    =
    R_{\rm peri}
    \left(
    \frac{M_{\rm dwarf}}
    {3M_{\rm MW}(<R_{\rm peri})}
    \right)^{1/3},
\end{equation}
where $r_{\rm J}$ is the Jacobi tidal radius, $R_{\rm peri}$ is the pericentric distance of the dwarf galaxy, $M_{\rm dwarf}$ is its adopted dynamical mass, and $M_{\rm MW}(<R_{\rm peri})$ is the MW mass enclosed within $R_{\rm peri}$.
The orbits of dSph galaxies are generally eccentric rather than circular. 
However, \citet{1962AJ.....67..471K} argued that the tidal radius of a satellite on a non-circular orbit can be approximated by evaluating the circular-orbit expression at pericenter. 
We therefore calculate $r_{\rm J}$ using $R_{\rm peri}$.

For the MW dark matter halo, we adopt the NFW density profile \citep{1997ApJ...490..493N}.
The corresponding enclosed mass within a Galactocentric radius $R$ is
\begin{equation}
    \label{eq:NFW}
    M_{\rm MW}(<R) =
    M_{200}
    \frac{\ln(1+x)-x/(1+x)}
    {\ln(1+c)-c/(1+c)},
\end{equation}
where $x=R/r_s$, and $r_s=r_{200}/c$. $M_{200}$ is the virial mass, and $c$ is the concentration parameter.
Here, $r_{200}$ is defined as the radius within which the mean enclosed density is $200\rho_{\rm crit}$ \citep{1997ApJ...490..493N}, where $\rho_{\rm crit}=3H_0^2/(8\pi G)$ \citep{1999astro.ph..5116H}. 
This definition gives $M_{200}=(4\pi/3)\,200\rho_{\rm crit}r_{200}^{3}$ and therefore $r_{200}=[3M_{200}/(4\pi\,200\rho_{\rm crit})]^{1/3}$.
We adopt the MW halo parameters from \citet{2020ApJ...894...10L}, who inferred $M_{200}=1.23^{+0.21}_{-0.18}\times10^{12}\,M_\odot$ and $c=9.4^{+2.8}_{-2.1}$ from the phase-space distribution of MW satellite galaxies.
Using Equation~(\ref{eq:NFW}), we estimate the enclosed MW mass at the pericentric distance of Draco, $R_{\rm peri}=58^{+11.4}_{-9.5}$ kpc \citep{2022ApJ...940..136P}.

We adopt the dynamical mass within half-light radius, $\log [M_{\rm dyn}(<r_{1/2})/M_\odot]=7.15\pm0.03$, \citep{2022NatAs...6..659B} as a proxy for the satellite mass $(M_{\rm dwarf})$.
The velocity-dispersion measurements used to estimate $M_{\rm dyn}(<r_{1/2})$ do not completely exclude the possible contribution from binary stars.
However, previous studies have found that binaries have only a minor effect on the measured velocity dispersion of Draco \citep{2024ApJ...970....1V}.
Moreover, since the Jacobi radius depends only on the cube root of the adopted dwarf-galaxy mass, plausible changes in the dynamical mass due to binaries would have a relatively small effect on $r_{\rm J}$ compared with the uncertainty associated with the adopted orbit.

The uncertainty range in $r_{\rm J}$ is estimated by evaluating all combinations of the upper and lower bounds of $M_{200}$, $c$, $R_{\rm peri}$, and $M_{\rm dyn}(<r_{1/2})$.
The minimum and maximum values obtained from this corner search are adopted as the lower and upper uncertainties, respectively.
From Equation~(\ref{eq:jacobi_radius}), we finally estimate the Jacobi tidal radius of Draco to be $r_{\rm J}=1.26^{+0.30}_{-0.25}\,\mathrm{kpc}$, corresponding to $57\arcmin.2^{+13.5}_{-11.1}$.
For comparison, adopting the pericentric distance of $r_{\rm peri}=100.0^{+21.9}_{-18.8}\,\mathrm{kpc}$ from \citet{2022NatAs...6..659B} gives a larger Jacobi radius of $r_{\rm J}=1.87^{+0.47}_{-0.39}\,\mathrm{kpc}$ ($84\farcm3^{+21.2}_{-17.8}$).

These estimates rely on several simplifying assumptions. 
We approximate the eccentric orbit of Draco by evaluating the circular-orbit expression at pericenter and model the MW as a static, spherically symmetric NFW halo, thereby neglecting its non-spherical structure \citep{2021MNRAS.501.2279V}, baryonic components, and time evolution \citep{2018Natur.563...85H}. 
We also treat $M_{\rm dyn}(<r_{1/2})$ as a single effective satellite mass and do not model the extended mass profile or non-spherical structure of Draco.

The quoted uncertainties are based on the minimum and maximum values from the corner search and are not formal confidence intervals. 
Therefore, the derived $r_{\rm J}$ should be regarded as a characteristic tidal scale rather than a sharply defined tidal boundary.
Variations in the adopted orbital and mass parameters, as well as the simplifying assumptions in the calculation, may shift the actual tidal boundary within a range comparable to the uncertainty.
Nevertheless, $r_{\rm J}$ provides a useful reference for assessing where the stellar distribution of Draco may become susceptible to the MW tidal field. 
We show the $r_{\rm J}$ as a circle in Figure~\ref{fig:contour}, since the calculation assumes spherical symmetry of dark-matter profile of dwarf galaxies.
This estimate relies on several simplifying assumptions. 
We approximate the eccentric orbit of Draco by evaluating the circular-orbit expression at pericenter and model the MW as a static, spherically symmetric NFW halo, thereby neglecting its non-spherical structure \citep{2021MNRAS.501.2279V}, baryonic components, and time evolution \citep{2018Natur.563...85H}. 
We also treat $M_{\rm dyn}(<r_{1/2})$ as a single effective satellite mass and do not model the extended mass profile or non-spherical structure of Draco.
The quoted uncertainties are based on the minimum and maximum values from the corner search and are not formal confidence intervals. 
Therefore, the derived $r_{\rm J}$ should be regarded as a characteristic tidal scale rather than a sharply defined tidal boundary.
Variations in the adopted orbital and mass parameters, as well as the simplifying assumptions in the calculation, may shift the actual tidal boundary within a range comparable to the uncertainty.
Nevertheless, $r_{\rm J}$ provides a useful reference for assessing where the stellar distribution of Draco may become susceptible to the MW tidal field. 
We show the $r_{\rm J}$ as a circle in Figure~\ref{fig:contour}, since the calculation assumes spherical symmetry of dark-matter profile of dwarf galaxies.

A comparison between the density contours and the predicted Jacobi radius shows that most of Draco's stellar distribution remains confined within the expected tidal boundary, suggesting that the observed stellar component is not strongly tidally truncated at the radius probed here (see Figure \ref{fig:contour}).
This is broadly consistent with the nearly flat velocity-dispersion profile reported by \citet{2005ApJ...631L.137M}.

A weak elongation along the major axis, extending preferentially toward the east, is visible in the contours below the $20\sigma$ level.
This extension lies approximately along the orbital direction, shown by the orange dashed line in Figure~\ref{fig:contour}.
At the lowest density level, the $5\sigma$ contour is more extended toward the northern side than toward the southern side.
\citet{2025ApJ...994..134D} identified eight high-probability Draco member RGB stars beyond the King tidal radius ($48\arcmin.1$; \citealp{2018ApJ...860...66M}) using heliocentric radial velocities and metallicities measured by DESI, together with proper motions from Gaia DR3.
Their positions, shown by the orange circles in the upper panels of Figure~\ref{fig:contour}, are preferentially distributed in the same general direction as this low-density extension (with one additional member star located to the east, approximately along the orbital direction, but outside these plots).
The spatial agreement between these independently selected stars and our density contours suggests the possible presence of a weak extended stellar distribution around Draco in this direction. 
Such a feature could arise from a past merger event, an interaction with a dark matter sub-halo \citep{2025MNRAS.tmp.1898V}, and/or tidal stripping by the Milky Way.

\subsection{Morphology of Bo\"otes I dSph}\label{sec:morphologyBootes}
In contrast, Bo\"otes I is often discussed as a dynamically disrupted dwarf galaxy \citep{2008MNRAS.385.1095F}. 
\citet{2021ApJ...923..218F} suggested that Bo\"otes I exhibits an extended stellar distribution reaching out to nearly $10\,r_{\rm h}$, based on the spatial distribution of BHB candidates. 
Furthermore, \citet{2016MNRAS.461.3702R} reported a weak S-shaped stellar structure around Bo\"otes~I that may be associated with tidal tails.
Our derived ellipticity is consistent with the previously reported value of $\epsilon=0.39\pm0.06$ \citep{2008ApJ...684.1075M} within the uncertainties.

The bottom panels of Figure~\ref{fig:contour} show the spatial distribution and the surface density map of Bo\"otes I.
The orange dashed line and arrow indicate the orbital direction, while the yellow arrow points toward the direction of the MW center.
The red circle indicates the Jacobi tidal radius (see Section~\ref{sec:morphologyDraco}), assuming a satellite mass $(m_{{\rm dwarf}})$ as dynamical mass within half-light radius, $\log [M_{\rm dyn}(<r_{1/2})/M_\odot] = 6.06^{+0.22}_{-0.30} M_\odot$, \citep{2022NatAs...6..659B} and a pericentric distance of $r_{\rm peri} = 37.9^{+7.5}_{-6.8}\,\rm{kpc}$ \citep{2022ApJ...940..136P}. 
The velocity dispersion used for this dynamical-mass estimate is based on \citet{2011ApJ...736..146K}, used repeated radial-velocity measurements to identify and exclude probable binary stars from their kinematic analysis.
We therefore expect any remaining binary contamination to have only a minor effect on the adopted dynamical mass.
From Equation~(\ref{eq:jacobi_radius}), we estimate the Jacobi tidal radius of Bo\"otes I to be $r_{\rm J}=0.41^{+0.18}_{-0.15}\,\mathrm{kpc}$, corresponding to $21\arcmin.6^{+9.3}_{-7.6}$ at the distance of Bo\"otes I.
When we instead adopt the pericentric distance of $r_{\rm peri} = 47.6^{+8.0}_{-8.7}\,\rm{kpc}$ from \citet{2022NatAs...6..659B}, the resulting Jacobi radius is consistent with our estimate within the uncertainties.

A S-shaped structure is visible at a level of $15\sigma$ above the background in the surface density map, extending along the orbital direction.
The spatial distribution of the selected MS stars, shown in the bottom-left panel of Figure~\ref{fig:contour}, clearly reveals the S-shaped tidal tails extending along the northern major axis.
Notably, the transition to the S-shaped morphology approximately coincides with the predicted Jacobi tidal radius.
We also find that the position angles of the isodensity contours differ between the inner and outer regions relative to the Jacobi tidal radius.
We discuss this point in Section \ref{sec:radial_ellipticity}.

\citet{2026ApJ...998...47S} reported a velocity gradient aligned with the orbital direction of Bo\"otes I and suggested that this gradient is related to tidal effects.
Considering the S-shaped morphology together with the reported velocity gradient, it is reasonable to consider that a tidal origin is the most plausible explanation for the structure observed in Bo\"otes I.

Furthermore, the existence of candidate BHB members beyond $\sim10\,r_{\rm h}$ \citep{2021ApJ...923..218F} suggests that Bo\"otes I may host extended stellar streams aligned with its orbit.
The future follow-up observation covering a wider area will be crucial to confirm its presence and extent of such tidal features.

We also find a weak symmetric extension along the direction perpendicular to the orbital direction, detected at a level of $15\sigma$ above the background.
As shown in the bottom-left panel of Figure \ref{fig:contour}, the western extension is affected by two bright foreground stars located at
$(\mathrm{R.A.},\ \mathrm{Decl.}) = (13^\mathrm{h}58^\mathrm{m}50\fs62,\ \mathrm{+14^{\circ}33\arcmin46\farcs58})$ and $(13^\mathrm{h}58^\mathrm{m}39\fs84,\ \mathrm{+14^{\circ}38\arcmin56\farcs44})$, with Gaia magnitudes of $G=7.1$ and $5.5$ mag, respectively.
The incompleteness caused by these bright stars may artificially produce a pinched or distorted morphology in this region.
Although an extended structure appears to be present toward the west, we cannot determine whether it is intrinsically narrow or broad.
In addition, toward the south-west, the clumpy structures are confirmed by the $15\sigma$ contour.

To investigate the nature of these possible sub-structures, which are visible in the contour plot, we compare their spatial distribution with that of Bo\"otes~I member candidates selected from {\it Gaia} DR3 (see Appendix~\ref{app:gaia_selection}).
The {\it Gaia}-selected stars show a distribution broadly aligned with the S-shaped structure seen in the HSC-selected MS sample (see Figure \ref{fig:Gaia_contour}).
In addition, a few member candidates overlap with the distorted western extension and with the southwestern and western clumps.
Although the number of stars is small, this spatial correspondence suggests that these features may be associated with tidal debris from Bo\"otes~I.

\section{Radial Profiles}\label{sec:radial_profile}
\subsection{Contamination Correction}
As mentioned in Section~\ref{sec:data}, we use HSC-SSP PDR3 Wide fields at similar Galactic latitudes of Draco and Bo\"otes I as control fields to estimate the foreground and background contamination. 
The CMDs of the control fields are shown in the right panels of Figure~\ref{fig:CMD}.

To construct the contamination profile, we first define red rectangles in the CMDs that are dominated by foreground and background
contaminants.
We compute the radial number-density profile of the objects within the red rectangle in the science field and normalize it to obtain the radial template profile of contamination, $T_{\rm sci}(r)$.
Following \citet{2025ApJ...993L...7S}, we assume that the number of contaminants selected within the red rectangle is proportional to that within the blue MS-selection polygon.

The contamination-corrected radial number density profile is then calculated as
\begin{equation}
\Sigma_{\rm dSph}(r)
=
\Sigma_{\rm obs}(r)
-
\Sigma_{\rm bg}(r),
\label{eq:surface_density}
\end{equation}
where $\Sigma_{\rm dSph}(r)$ is the radial number density profile of each dSph, and $\Sigma_{\rm obs}(r)$ is the radial number density profiles of the MS candidate stars selected within the blue polygon in Figure \ref{fig:CMD}.
$\Sigma_{\rm bg}(r)$ is defined as
\begin{equation}
\Sigma_{\rm bg}(r)
=
\left(
\frac{N_{\rm sci}^{\rm red}}
     {N_{\rm ctrl}^{\rm red}}
\right)
N_{\rm ctrl}^{\rm blue}
T_{\rm sci}(r),
\label{eq:background}
\end{equation}
where $N_{\rm sci}^{\rm red}$ and $N_{\rm ctrl}^{\rm red}$ are the weighted stellar numbers within the red rectangle in the science and control fields, respectively, $N_{\rm ctrl}^{\rm blue}$ is the number of control-field stars within the blue MS-selection polygon, and $T_{\rm sci}(r)$ is the normalized template profiles of contamination derived from the science field.
The ratio $N_{\rm sci}^{\rm red}/N_{\rm ctrl}^{\rm red}$ scales the number of stars within the blue polygon of the control field to the expected number of contaminants within the corresponding selection region of the science field.
Multiplying this estimated number by $T_{\rm sci}(r)$ yields the radial number density profile of the contamination ($\Sigma_{\rm bg}(r)$), which is subtracted from the observed profile ($\Sigma_{\rm obs}(r)$).
Through this procedure, we obtain radial number density profiles with a reduced contribution from foreground and background sources.
The resulting profiles are shown in Figures~\ref{fig:Dra_profile} and \ref{fig:Boo1_profile}.
\subsection{Radial Profiles of Draco}\label{sec:profiledraco}
    \begin{figure*}[ht!]
        \begin{center}
        \includegraphics[width=16cm]{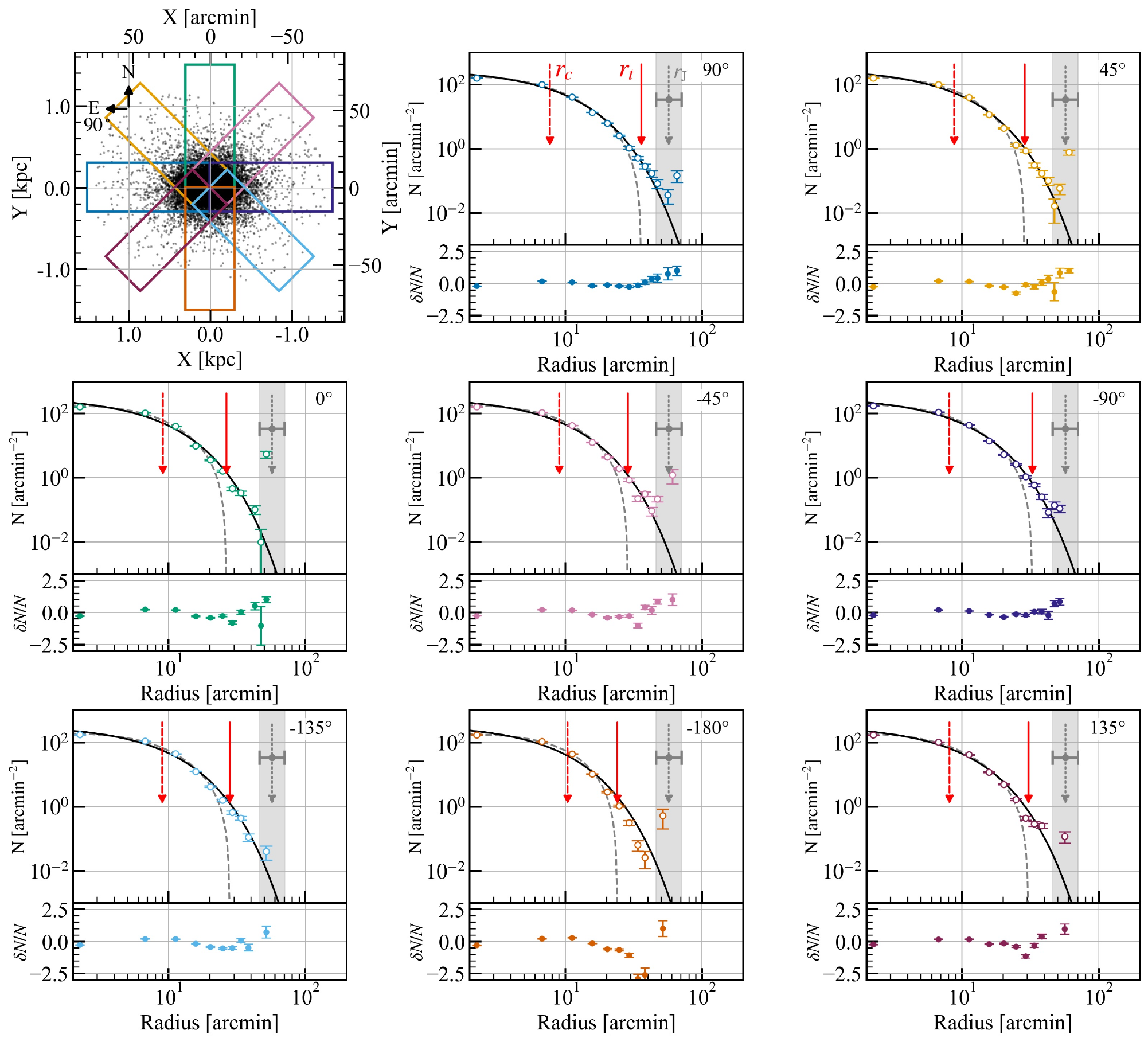}
        \end{center}
        \caption
        {
        The top-left panel shows the spatial distribution of the selected MS stars, with eight regions defined by $-45^\circ$ azimuthal intervals, starting from North ($0^\circ$) to East ($90^\circ$).
        The remaining panels present, except for the top-left panel showing the spatial distribution, present the observed radial number density profiles together with the best-fitting exponential model (black solid line) and King model (gray dashed line; \citealp{1962AJ.....67..471K}).
        In this analysis, the radial profiles are constructed using projected circular coordinates rather than elliptical radii. 
        The red solid and dashed arrows indicate the core radius ($r_c$) and the tidal radius ($r_t$), respectively.
        The gray dotted arrows indicate the location of Jacobi tidal radius, while the gray shaded region shows its uncertainty range.
        The uncertainties consider the errors of pericentric distance, dynamical mass, and assumed MW potential parameters.
        The lower sub-panel in each panel shows the residuals with respect to the exponential profile.
        }\label{fig:Dra_profile}
    \end{figure*}
To investigate whether Draco exhibits extended structures, we derive radial number density profiles in eight directions.
We use projected circular radii rather than elliptical radii, because the outer stellar structures may not follow the inner elliptical morphology.
For this assessment, we define rectangular regions over the range $-180^\circ \leq \theta < 180^\circ$ at $45^\circ$ intervals, adopting fixed dimensions of 600 pc for the short side and 1500 pc for the long side.
The top-left panel of Figure \ref{fig:Dra_profile} shows the corresponding rectangles overlaid on the spatial distribution of MS stars, while the other panels present the radial number density profiles for the eight position angles.
We then fit exponential and King profiles \citep{1962AJ.....67..471K} to the profiles in each direction.
The derived core and tidal radii are indicated in Figure \ref{fig:Dra_profile} as red dashed and solid arrows, respectively.
By comparing the exponential model with the observed profiles, we show the residuals in the bottom subpanels.

We do not find a significant excess over the single exponential and King profiles in most directions.
It should be noted that at larger radii, beyond $R\sim60\arcmin$, the stellar counts become very small, and the corresponding regions lie close to the edges of the imaging footprint, making the measurements increasingly sensitive to small-number statistics and uncertainties in the effective survey area.
Nevertheless, slight excesses are visible toward $90^\circ$ (east), $45^\circ$ (northeast), $-45^\circ$ (northwest), and $-90^\circ$ (west).
These directions are broadly consistent with the distribution of the extra-tidal stars identified by \citet{2025ApJ...994..134D}, shown as orange circles in Figure~\ref{fig:contour}.

We also fit double-exponential profiles to the radial profiles.
Based on the Bayesian Information Criterion (BIC), the single-exponential model is preferred in all directions.
Therefore, we do not find statistically significant evidence for an additional extended stellar component in Draco from the present radial-profile analysis.
A more sensitive test will require secure member identification among the faint MS stars, ideally through proper-motion measurements extending to these magnitudes.
\subsection{Radial Profiles of Bo\"otes I}
    \begin{figure*}[ht!]
        \begin{center}
        \includegraphics[width=16cm]{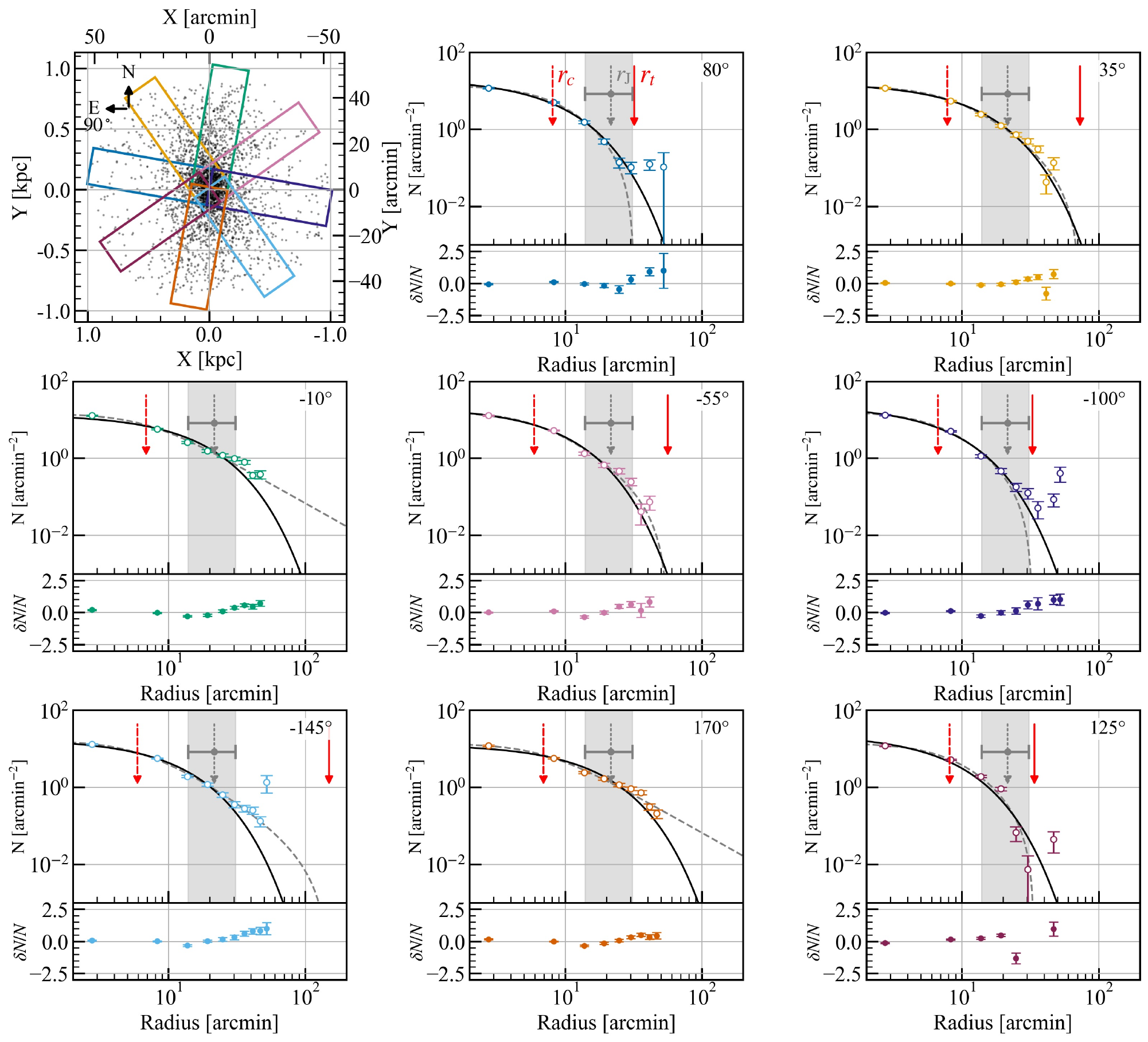}
        \end{center}
        \caption
        {
        The same as Figure \ref{fig:Dra_profile}, but for the Bo\"{o}tes~I dSph, with azimuthal regions starting from $80^\circ$.
        }\label{fig:Boo1_profile}
    \end{figure*}
As shown in Section \ref{sec:morphologyBootes}, Bo\"otes I has the asymmetric S-shape stellar distribution.
To quantify the azimuthal variation of this excess relative to a simple single-exponential profile, we construct radial number-density profiles in eight different directions, as done for Draco.
The aim is slightly different; we investigate potential azimuthal variations and asymmetries in the stellar distribution.
The rectangular regions along the position angles of $-10^\circ$, which is the direction of the S-shaped structure, and spaced at $45^\circ$ intervals.
For Bo\"otes~I, we adopt narrower rectangular regions ($300\times1000$ pc) than those used for Draco ($600\times1500$ pc).
This choice is motivated by the relatively narrow S-shaped extension of Bo\"otes~I; a broader region would include additional irregular substructures and could contaminate its directional signature.
These angles are chosen to cover the direction of the S-shaped structure.
The top-left panel of Figure \ref{fig:Boo1_profile} shows the adopted rectangles, while the remaining panels present the corresponding radial number density profiles.
We fit exponential and King profiles \citep{1962AJ.....67..471K} in each direction and show the residuals relative to the exponential model in the bottom subpanels.

We first focus on the position angles of $170^\circ$ and $-10^\circ$, which correspond to the S-shaped structure identified in Figure~\ref{fig:contour}.
The radial profiles in these directions deviate from the best-fitting single exponential profile, and the King-model fits also yield larger tidal radii.
Excesses over the single exponential model are also seen at position angles of $80^\circ$, $35^\circ$, $-55^\circ$, $-100^\circ$, and $-145^\circ$.
To further assess these deviations, we fit double-exponential models to the radial profiles and compare them with the single-exponential fits using the BIC.
We define $\Delta{\rm BIC}={\rm BIC}_{\rm double}-{\rm BIC}_{\rm single}$, such that negative values indicate a preference for the double-exponential model.
We find at position angles of $35^\circ$, $-10^\circ$, $-55^\circ$, $-145^\circ$, and $170^\circ$, with $\Delta{\rm BIC}\lesssim-10$ in these directions, strongly favoring the double-exponential profile over the single-exponential profile.
These directions broadly correspond to additional distortions visible in the low-density contours in Figure~\ref{fig:contour}, suggesting that the outer stellar distribution of Bo\"otes~I is extended in several directions rather than being confined solely to the main S-shaped feature.

The gray dotted arrows indicate the location of the Jacobi tidal radius calculated in Section~\ref{sec:morphologyBootes}.
Notably, in several azimuthal directions, the observed profiles begin to exceed the best-fitting single exponential profile at $r \sim 20\arcmin$, close to the estimated Jacobi tidal radius.
A similar break at approximately $20\arcmin$ was reported for Bo\"otes~I by \citet{2016MNRAS.461.3702R}.
This correspondence suggests that the extended stellar component may be related to tidal stripping beyond the Jacobi tidal radius \citep{2008ApJ...673..226P}.

\subsection{The Radial Variation of the Structure Parameters}\label{sec:radial_ellipticity}
    \begin{figure*}[ht!]
        \begin{center}
        \includegraphics[width=16cm]{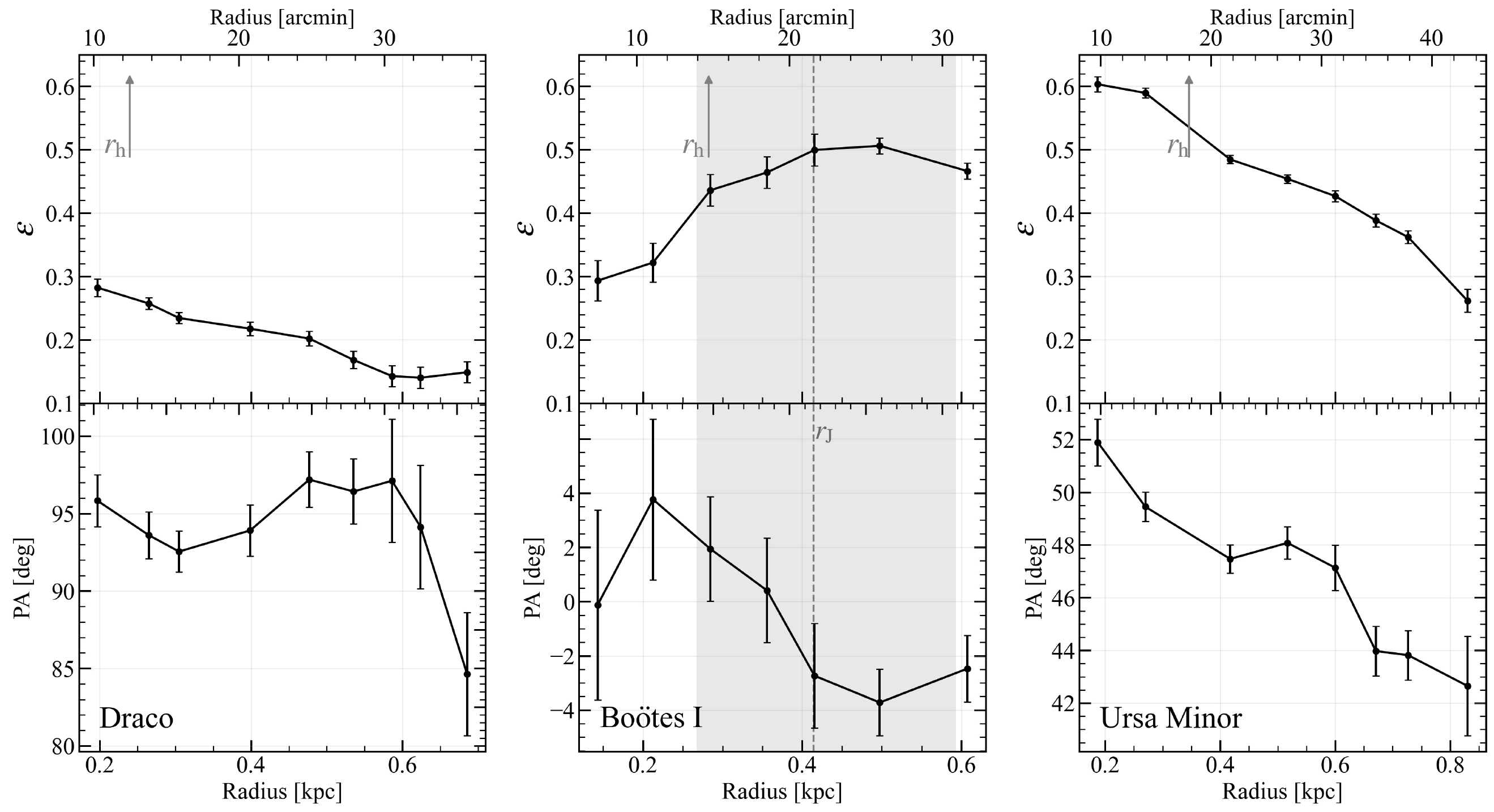}
        \end{center}
        \caption
            {
            Radial profiles of ellipticity (top panels) and position angle (bottom panels) for the Draco, Bo\"otes I, and Ursa Minor dwarf spheroidal galaxies.
            The data points are derived from ellipse fitting to isodensity contours at fixed significance levels (in units of $\sigma$) in the stellar surface density maps.
            The radius is the circularized radius, $R_{\rm circ}=\sqrt{A/\pi}=\sqrt{ab}$, where $A$ is the area enclosed by each isodensity contour.
            We set $10, 15, 20, 30, 50, 100, 300, 500$, and $1000 \sigma$ for Draco, $10, 15, 20, 30, 50$, and $100 \sigma$ for Bo\"otes I, and $10, 15, 20, 30, 50, 100, 300$, and $500 \sigma$ for Ursa Minor.
            The uncertainties are estimated from 100 Poisson realizations of the stellar surface-density maps, generated by independently Poisson-sampling each pixel using its original value as the expectation value, for which the same contour selection and moment-based shape measurements are repeated.
            The half-light radius of each dSph is indicated by a gray arrow.
        For Bo\"otes I, the Jacobi tidal radius is shown as a dashed gray line.
                    }\label{fig:radial_ellipticity}
    \end{figure*}
To assess how the morphology change with the distance from the center, we investigate the radial profiles of ellipticity and position angle by fitting ellipses to the isodensity contours for Draco, Bo\"otes I, and Ursa Minor.
We include Ursa Minor as a comparison system because it has metallicity, orbital properties, and stellar mass similar to those of Draco.
The data of Ursa Minor are based on the Subaru/HSC observations presented in \citet{2025ApJ...993L...7S}.
This data has a comparable depth of photometry and spatial coverage to the data of this paper, thus we can directly compare to the results of Draco and Bo\"otes I.
For the fitting procedure, we adopt the same smoothing scale and binning scheme as used in Figure \ref{fig:contour}.

To estimate the ellipticity and position angle at each isodensity, we compute the second-older moments of the largest connected region enclosed by the corresponding isodensity contour. 
The radius is the circularized radius, $R_{\rm circ}=\sqrt{A/\pi}=\sqrt{ab}$, where $A$ is the area enclosed by each isodensity contour.
The ellipse parameters are then derived from the eigenvalues and eigenvectors of the moment following the \citet{1980SPIE..264..208S}.
The second-order moments are derived as follows:
\begin{equation}
I_{xx} = \frac{1}{N} \sum_{i=1}^{N} (x_i - x_0)^2,
\end{equation}

\begin{equation}
I_{yy} = \frac{1}{N} \sum_{i=1}^{N} (y_i - y_0)^2,
\end{equation}

\begin{equation}
I_{xy} = \frac{1}{N} \sum_{i=1}^{N} (x_i - x_0)(y_i - y_0),
\end{equation}
where $(x_0, y_0)$ denotes the centroid of the selected region and $N$ is the number of pixels enclosed by the isodensity contour.
The position angle $\theta$ of the ellipse is then derived from
\begin{equation}
\tan(2\theta) = \frac{2 I_{xy}}{I_{xx} - I_{yy}}.
\end{equation}
The semi-major and semi-minor axes ($a$ and $b$) are obtained from the eigenvalues of the moment matrix, which are equivalent to the analytical expressions given in \citet{1980SPIE..264..208S}.
\begin{equation}
a^2
=
\frac{I_{xx}+I_{yy}}{2}
+
\frac{1}{2}
\sqrt{(I_{xx}-I_{yy})^2 + 4I_{xy}^2},
\end{equation}
\begin{equation}
b^2
=
\frac{I_{xx}+I_{yy}}{2}
-
\frac{1}{2}
\sqrt{(I_{xx}-I_{yy})^2 + 4I_{xy}^2}.
\end{equation}
The ellipticity is defined as $\epsilon=1-\frac{b}{a}$.

The resultant radial ellipticity and position angle profiles are shown in Figure \ref{fig:radial_ellipticity}.
A comparison of the ellipticity profiles reveals that both Draco and Ursa Minor show decreasing ellipticity at larger radii, suggesting a transition toward more circular morphologies in their outskirts.
In this study, we find no clear signature of an extended stellar distribution in Draco.
In addition, the estimated Jacobi tidal radius is larger than the observed extent of the stellar distribution.
These results may suggest that Draco has experienced weaker tidal effects from MW than the other dSphs considered in this study. 
For Ursa Minor, \citet{2025ApJ...993L...7S} identified an extended distribution of main-sequence stars preferentially along the minor-axis direction.
In our analysis, the ellipticity decreases toward larger radii and reaches $\epsilon \approx 0.26$ in the outskirts.
This decrease may be explained by the increasing contribution of the minor-axis stellar component, which makes the outer morphology appear rounder than the inner distribution.

In contrast, Bo\"otes I exhibit an increase in ellipticity with radius.
This increasing trend is similar to that of the Ursa Major II case \citep{2010AJ....140..138M}. 
\citet{2002AJ....124..127J} demonstrated through N-body simulations that tidally disturbed dwarf galaxies generally exhibit increasing ellipticity with radius.
Considering the increasing ellipticity, Bo\"otes I could be affected by the tidal disturbance.

Regarding the position-angle profiles, Ursa Minor shows a gradual change in position angle with increasing distance from the center, whereas the profiles of Draco and Bo\"otes~I are broadly consistent with being constant over most of the radial range within the uncertainties.
Although the outermost bin of Draco shows a sudden change in position angle, it does not exhibit the gradual radial trend seen in Ursa Minor.
Moreover, the ellipticity of Draco decreases to $\epsilon \sim 0.15$ in the outer region, making the morphology close to circular.
Thus, the apparent change in the outermost bin may not represent a significant morphological twist.

Ursa Minor shows both a systematic position-angle variation and the strongest radial change in ellipticity among the dSphs investigate in this study. 
In a triaxial system, radial changes in intrinsic shape can also produce apparent variations in position angle through projection effects \citep{1980MNRAS.193..885B}. 
Thus, the observed twist in Ursa Minor may partly reflect its strong ellipticity gradient rather than an intrinsic twist of the stellar distribution.

\section{Discussions}\label{sec:discussion}
Based on homogeneous Subaru/HSC imagery, the outskirts of Draco and Bo\"otes I dSphs exhibit distinct properties.  Here we discuss the properties of each galaxy and its origin, together with the additional sample, Ursa Minor dSph taken from \citet{2025ApJ...993L...7S}.
\subsection{Draco and Ursa Minor}
In this study, we demonstrate that Draco does not exhibit a prominent extended stellar distribution.
Weak extensions are detected toward the east and northwest directions, but the double-exponential profile is not preferred over the single-exponential one.
Its ellipticity decreases toward larger radii, while the increasing trend of Bo\"otes I.
The outer stellar distribution of Draco appears relatively less distorted compared with those of Bo\"otes~I and Ursa Minor.

Ursa Minor and Draco have comparable stellar masses ($M_{\star,\rm UMi}=0.29\times10^{6}\,M_\odot$ and $M_{\star,\rm Dra}=0.29\times10^{6}\,M_\odot$; \citealp{2012AJ....144....4M}) and broadly similar orbital properties around the MW (Distance of Pericenter [kpc]; $R_{\rm Peri, UMi}=41.8^{+5.3}_{-4.5},\ R_{\rm Peri, Dra}=40.4^{+6.5}_{-5.4}$, orbital eccentricity; $e_{\rm UMi}=0.29\pm0.03,\ e_{\rm Dra}=0.30\pm0.04$).
Nevertheless, substantial differences have been identified in their internal stellar populations and spatial structures.  
Ursa Minor contains multiple chemodynamical stellar populations and exhibits a spatially extended stellar distribution \citep{2020MNRAS.495.3022P,2025ApJ...993L...7S}, whereas Draco shows a comparatively regular elliptical morphology.
A direct comparison between these otherwise similar systems may therefore provide important clues to how differences in their formation and evolutionary histories produce distinct present-day structures.

To perform this comparison, we estimate the Jacobi tidal radius of Ursa Minor by adopting a dynamical mass of $\log M_{\rm dyn}(r_{1/2})=7.21\pm0.03$ \citep{2022NatAs...6..659B} and a pericentric distance of $R_{\rm peri}=55.7^{+8.4}_{-7.0}\,\mathrm{kpc}$ \citep{2022ApJ...940..136P}.
The resulting Jacobi tidal radius is $r_{\rm J}=1.29^{+0.26}_{-0.22}\,\mathrm{kpc}$, which is consistent within the uncertainties with that estimated for Draco, $r_{\rm J}=1.26^{+0.30}_{-0.25}\,\mathrm{kpc}$.
However, the half-light radius of Ursa Minor ($345.8$ pc; \citealp{2025PASJ...77.1259S}) is larger than that of Draco ($239.6$ pc; this work), indicating that the stellar distribution of Ursa Minor is more spatially extended than that of Draco.
The compact stellar distribution of Draco may indicate that its stars are more tightly bound within its dark matter potential, making the system less affected by MW tidal effects.
\citet{2026arXiv260424855P} compared the dark matter density profiles of Draco and Ursa Minor using multi-component axisymmetric dynamical models based on stellar distribution functions.
They modeled each galaxy with two chemodynamically distinct stellar populations embedded in a spherical dark matter potential and inferred a cuspy inner dark matter profile for Draco ($\gamma=0.98^{+0.28}_{-0.26}$), but a shallower, more cored profile for Ursa Minor ($\gamma=0.37^{+0.31}_{-0.24}$).
The inferred dark matter distribution of Draco is more centrally concentrated than that of Ursa Minor, with a dark matter density at $r=150$ pc approximately five times higher (see Table C.3 of \citealp{2026arXiv260424855P}).
The higher central dark matter density in Draco may generate a stronger restoring force and keep its inner stellar component more tightly bound \citep{2015MNRAS.449L..46E}.
Although the response of the outer stellar distribution also depends on the orbital history of dwarf galaxies, the more centrally concentrated dark matter halo of Draco may contribute to its comparatively regular morphology.

\subsection{Bo\"otes~I}
Bo\"otes I has a stellar mass approximately an order of magnitude lower than those of Ursa Minor and Draco ($M_{\star,\mathrm{\text{Bo\"otes,I}}} = 0.029\times10^{6}\,M_\odot$; \citealp{2012AJ....144....4M}). 
Despite its much lower stellar mass, its half-light radius, $r_h = 282.79$ pc, lies between those of Draco and Ursa Minor. 
In this study we identify an extended S-shaped stellar structure that may be associated with tidal tails, suggesting that Bo\"otes I is unusually extended for its stellar mass. 
Such a diffuse stellar distribution may reflect, at least in part, differences in the underlying dark-matter density profile and its susceptibility to tidal perturbations.
Recently, \citet{2026arXiv260626218N} inferred a low inner dark-matter density for Bo\"otes I, $\rho_{150}=0.36^{+0.15}_{-0.11}\times10^{8}\,M_\odot\,\mathrm{kpc}^{-3}$, placing it among the lowest-density dwarf galaxies at comparable stellar masses.
They also inferred a shallow logarithmic density slope, $\gamma_{\rm in}=-0.43^{+0.49}_{-0.42}$ at $0.015$ of the virial radius, consistent with a relatively shallow or cored inner profile, although the inner slope remains poorly constrained.
The contrasting outer morphologies of these systems may therefore reflect not only differences in their orbital histories, but also differences in the internal structure of their dark-matter halos. 
In particular, the relatively low inner dark-matter density of Bo\"otes I provides weaker gravitational binding for its stellar component, making it more susceptible to tidal perturbations and contributing to the extended, distorted morphology identified in this work.

\section{Conclusions}\label{sec:concl}
In this work, we investigated the detailed stellar distributions of the Draco and Bo\"otes~I, using old MS stars detected in the deep, wide-field HSC dataset.
We derived their global structural parameters and examined their outer morphologies through stellar surface-density maps, directional radial number-density profiles, and radial variations in ellipticity and position angle.

The derived half-light radii are $r_h = 239.6^{+1.9}_{-2.0}$ pc for Draco and $r_h = 282.79^{+5.84}_{-6.02}$ pc for Bo\"otes~I.
Draco has a relatively round and compact stellar distribution, with a global ellipticity of $\epsilon = 0.266 \pm 0.007$.
No prominent extended stellar component is detected, although weak extensions are present in the eastern and northwestern directions.
In contrast, Bo\"otes~I is more elongated, with $\epsilon = 0.428^{+0.014}_{-0.015}$, and exhibits a prominent S-shaped stellar structure extending approximately along its orbital direction.

The radial number-density profiles further reveal a clear difference between the two systems.
For Draco, weak extensions are visually seen toward the directions of previously identified extra tidal stars.
However, a single-exponential profile is preferred in all directions, with no statistically significant additional outer component.
For Bo\"otes~I, the stellar density begins to exceed the single-exponential profile at $r \sim 20\arcmin$ in several directions, close to its estimated Jacobi tidal radius, and double-exponential profiles are preferred along the directions of the S-shaped structure.
Together with its alignment with the orbital direction, these results suggest that the S-shaped structure may represent tidal tails produced by interactions with the Milky Way.

The radial ellipticity and position angle profiles also show contrasting behavior.
The ellipticity of Draco decreases toward larger radii, reaching $\epsilon \sim 0.15$ in the outskirts, indicating an increasingly circular outer morphology.
In contrast, the ellipticity of Bo\"otes~I systematically increases with radius, consistent with the behavior expected for a tidally disturbed stellar system.
The position angles of both galaxies remain broadly constant over most of the radial range within the uncertainties. 

These results demonstrate that dSphs with broadly similar Galactic environments can show substantially different responses in their outer stellar distributions.
In particular, the increasing ellipticity, density excess near the Jacobi radius, and S-shaped morphology of Bo\"otes~I consistently point toward tidal disturbance, whereas Draco remains comparatively compact and less distorted shape.
This diversity may reflect not only differences in orbital evolution, but also differences in the internal dark-matter structure that determines how strongly the stellar component is bound against the Milky Way tidal field.

While photometric morphology studies provide important clues, they alone are insufficient to determine the origin of extended stellar structures.
Constraining their formation mechanisms requires detailed dynamical information.
\citet{2017ApJ...850..144E} showed that prolate rotation in dwarf galaxies can arise as a consequence of dwarf-dwarf mergers, highlighting the possibility that merger-driven processes may also play a role in shaping extended stellar components.
On the other hand, \citet{2026ApJ...998...47S} noted that a velocity gradient aligned with the orbital direction can be a signature of tidal disruption.
Therefore, to fully understand the origin of the extended structures in these systems, it is necessary to investigate their detailed stellar dynamics even in the outskirts of dwarf.
The wide field of view of the Subaru Prime Focus Spectrograph (\text{\={O}nohi\textquoteleft ula}) provides a unique opportunity to carry out such observations \citep{2026arXiv260409875C}.

\begin{acknowledgments}
This work was supported by JSPS KAKENHI Grant Numbers JP18H05875, JP20K04031, JP20H05855, JP25K01047.
This work was supported by JST SPRING, Japan Grant Number JPMJSP2104.
The authors thank the referee for helpful comments on the manuscript.

The Hyper Suprime-Cam (HSC) collaboration includes the astronomical communities of Japan and Taiwan, and Princeton University. 
The HSC instrumentation and software were developed by the National Astronomical Observatory of Japan (NAOJ), the Kavli Institute for the Physics and Mathematics of the Universe (Kavli IPMU), the University of Tokyo, the High Energy Accelerator Research Organization (KEK), the Academia Sinica Institute for Astronomy and Astrophysics in Taiwan (ASIAA), and Princeton University. 
Data analysis was carried out on the large-scale data analysis system co-operated by the Astronomy Data Center (ADC) and Subaru Telescope, NAOJ. 
This research is based on data collected at the Subaru Telescope.
We are honored and grateful for the opportunity of observing the Universe from Maunakea, which has the cultural, historical, and natural significance in Hawaii.
The data are obtained from SMOKA, which is operated by the NAOJ/ADC.

The Pan-STARRS1 Surveys (PS1) have been made possible through contributions of the Institute for Astronomy, the University of Hawaii, the Pan-STARRS Project Office, the Max-Planck Society and its participating institutes, the Max Planck Institute for Astronomy, Heidelberg and the Max Planck Institute for Extraterrestrial Physics, Garching, The Johns Hopkins University, Durham University, the University of Edinburgh, Queen's University Belfast, the Harvard-Smithsonian Center for Astrophysics, the Las Cumbres Observatory Global Telescope Network Incorporated, the National Central University of Taiwan, the Space Telescope Science Institute, the National Aeronautics and Space Administration under Grant No. NNX08AR22G issued through the Planetary Science Division of the NASA Science Mission Directorate, the National Science Foundation under Grant No. AST-1238877, the University of Maryland, and Eotvos Lorand University (ELTE).

\end{acknowledgments}
\software{
Astropy \citep{2013A&A...558A..33A,2018AJ....156..123A,2022ApJ...935..167A},
NumPy \citep{2020Natur.585..357H},
SciPy \citep{2020NatMe..17..261V},
Matplotlib \citep{2007CSE.....9...90H},
pandas \citep{mckinney2010data},
emcee \citep{2013PASP..125..306F},
corner \citep{2016JOSS....1...24F},
HSC Pipeline \citep{2018PASJ...70S...5B}
}
\bibliography{ref_boo1_draco.bib}

\begin{thebibliography}{}
\expandafter\ifx\csname natexlab\endcsname\relax\def\natexlab#1{#1}\fi
\providecommand{\url}[1]{\href{#1}{#1}}
\providecommand{\dodoi}[1]{doi:~\href{http://doi.org/#1}{\nolinkurl{#1}}}
\providecommand{\doeprint}[1]{\href{http://ascl.net/#1}{\nolinkurl{http://ascl.net/#1}}}
\providecommand{\doarXiv}[1]{\href{https://arxiv.org/abs/#1}{\nolinkurl{https://arxiv.org/abs/#1}}}

\bibitem[{H. {Aihara} {et~al.}(2022){Aihara}, {AlSayyad}, {Ando}, {Armstrong}, {Bosch}, {Egami}, {Furusawa}, {Furusawa}, {Harasawa}, {Harikane}, {Hsieh}, {Ikeda}, {Ito}, {Iwata}, {Kodama}, {Koike}, {Kokubo}, {Komiyama}, {Li}, {Liang}, {Lin}, {Lupton}, {Lust}, {MacArthur}, {Mawatari}, {Mineo}, {Miyatake}, {Miyazaki}, {More}, {Morishima}, {Murayama}, {Nakajima}, {Nakata}, {Nishizawa}, {Oguri}, {Okabe}, {Okura}, {Ono}, {Osato}, {Ouchi}, {Pan}, {Plazas Malag{\'o}n}, {Price}, {Reed}, {Rykoff}, {Shibuya}, {Simunovic}, {Strauss}, {Sugimori}, {Suto}, {Suzuki}, {Takada}, {Takagi}, {Takata}, {Takita}, {Tanaka}, {Tang}, {Taranu}, {Terai}, {Toba}, {Turner}, {Uchiyama}, {Vijarnwannaluk}, {Waters}, {Yamada}, {Yamamoto}, \& {Yamashita}}]{2022PASJ...74..247A}
{Aihara}, H., {AlSayyad}, Y., {Ando}, M., {et~al.} 2022, \bibinfo{title}{{Third data release of the Hyper Suprime-Cam Subaru Strategic Program},} \pasj, 74, 247, \dodoi{10.1093/pasj/psab122}

\bibitem[{ {Astropy Collaboration} {et~al.}(2013){Astropy Collaboration}, {Robitaille}, {Tollerud}, {Greenfield}, {Droettboom}, {Bray}, {Aldcroft}, {Davis}, {Ginsburg}, {Price-Whelan}, {Kerzendorf}, {Conley}, {Crighton}, {Barbary}, {Muna}, {Ferguson}, {Grollier}, {Parikh}, {Nair}, {Unther}, {Deil}, {Woillez}, {Conseil}, {Kramer}, {Turner}, {Singer}, {Fox}, {Weaver}, {Zabalza}, {Edwards}, {Azalee Bostroem}, {Burke}, {Casey}, {Crawford}, {Dencheva}, {Ely}, {Jenness}, {Labrie}, {Lim}, {Pierfederici}, {Pontzen}, {Ptak}, {Refsdal}, {Servillat}, \& {Streicher}}]{2013A&A...558A..33A}
{Astropy Collaboration}, {Robitaille}, T.~P., {Tollerud}, E.~J., {et~al.} 2013, \bibinfo{title}{{Astropy: A community Python package for astronomy},} \aap, 558, A33, \dodoi{10.1051/0004-6361/201322068}

\bibitem[{ {Astropy Collaboration} {et~al.}(2018){Astropy Collaboration}, {Price-Whelan}, {Sip{\H{o}}cz}, {G{\"u}nther}, {Lim}, {Crawford}, {Conseil}, {Shupe}, {Craig}, {Dencheva}, {Ginsburg}, {VanderPlas}, {Bradley}, {P{\'e}rez-Su{\'a}rez}, {de Val-Borro}, {Aldcroft}, {Cruz}, {Robitaille}, {Tollerud}, \& {Astropy Contributors}}]{2018AJ....156..123A}
{Astropy Collaboration}, {Price-Whelan}, A.~M., {Sip{\H{o}}cz}, B.~M., {et~al.} 2018, \bibinfo{title}{{The Astropy Project: Building an Open-science Project and Status of the v2.0 Core Package},} \aj, 156, 123, \dodoi{10.3847/1538-3881/aabc4f}

\bibitem[{ {Astropy Collaboration} {et~al.}(2022){Astropy Collaboration}, {Price-Whelan}, {Lim}, {Earl}, {Starkman}, {Bradley}, {Shupe}, {Patil}, {Corrales}, {Brasseur}, {N{\"o}the}, {Donath}, {Tollerud}, {Morris}, {Ginsburg}, {Vaher}, {Weaver}, {Tocknell}, {Jamieson}, {van Kerkwijk}, {Robitaille}, {Merry}, {Bachetti}, {G{\"u}nther}, \& {Astropy Project Contributors}}]{2022ApJ...935..167A}
{Astropy Collaboration}, {Price-Whelan}, A.~M., {Lim}, P.~L., {et~al.} 2022, \bibinfo{title}{{The Astropy Project: Sustaining and Growing a Community-oriented Open-source Project and the Latest Major Release (v5.0) of the Core Package},} \apj, 935, 167, \dodoi{10.3847/1538-4357/ac7c74}

\bibitem[{G. {Battaglia} \& C. {Nipoti}(2022){Battaglia} \& {Nipoti}}]{2022NatAs...6..659B}
{Battaglia}, G., \& {Nipoti}, C. 2022, \bibinfo{title}{{Stellar dynamics and dark matter in Local Group dwarf galaxies},} Nature Astronomy, 6, 659, \dodoi{10.1038/s41550-022-01638-7}

\bibitem[{K. {Bekki} \& M. {Chiba}(2001){Bekki} \& {Chiba}}]{2001ApJ...558..666B}
{Bekki}, K., \& {Chiba}, M. 2001, \bibinfo{title}{{Formation of the Galactic Stellar Halo. I. Structure and Kinematics},} \apj, 558, 666, \dodoi{10.1086/322300}

\bibitem[{V. {Belokurov} {et~al.}(2006){Belokurov}, {Zucker}, {Evans}, {Wilkinson}, {Irwin}, {Hodgkin}, {Bramich}, {Irwin}, {Gilmore}, {Willman}, {Vidrih}, {Newberg}, {Wyse}, {Fellhauer}, {Hewett}, {Cole}, {Bell}, {Beers}, {Rockosi}, {Yanny}, {Grebel}, {Schneider}, {Lupton}, {Barentine}, {Brewington}, {Brinkmann}, {Harvanek}, {Kleinman}, {Krzesinski}, {Long}, {Nitta}, {Smith}, \& {Snedden}}]{2006ApJ...647L.111B}
{Belokurov}, V., {Zucker}, D.~B., {Evans}, N.~W., {et~al.} 2006, \bibinfo{title}{{A Faint New Milky Way Satellite in Bootes},} \apjl, 647, L111, \dodoi{10.1086/507324}

\bibitem[{L. {Benacchio} \& G. {Galletta}(1980){Benacchio} \& {Galletta}}]{1980MNRAS.193..885B}
{Benacchio}, L., \& {Galletta}, G. 1980, \bibinfo{title}{{Triaxiality in elliptical galaxies},} \mnras, 193, 885, \dodoi{10.1093/mnras/193.4.885}

\bibitem[{E. {Bertin}(2011){Bertin}}]{2011ASPC..442..435B}
{Bertin}, E. 2011, in Astronomical Society of the Pacific Conference Series, Vol. 442, Astronomical Data Analysis Software and Systems XX, ed. I.~N. {Evans}, A.~{Accomazzi}, D.~J. {Mink}, \& A.~H. {Rots}, 435

\bibitem[{E. {Bertin}(2013){Bertin}}]{2013ascl.soft01001B}
{Bertin}, E. 2013, \bibinfo{title}{{PSFEx: Point Spread Function Extractor},}, Astrophysics Source Code Library, record ascl:1301.001 \doeprint{1301.001}

\bibitem[{J. {Binney} \& S. {Tremaine}(2008){Binney} \& {Tremaine}}]{2008gady.book.....B}
{Binney}, J., \& {Tremaine}, S. 2008, {Galactic Dynamics: Second Edition}

\bibitem[{A.~Z. {Bonanos} {et~al.}(2004){Bonanos}, {Stanek}, {Szentgyorgyi}, {Sasselov}, \& {Bakos}}]{2004AJ....127..861B}
{Bonanos}, A.~Z., {Stanek}, K.~Z., {Szentgyorgyi}, A.~H., {Sasselov}, D.~D., \& {Bakos}, G.~{\'A}. 2004, \bibinfo{title}{{The RR Lyrae Distance to the Draco Dwarf Spheroidal Galaxy},} \aj, 127, 861, \dodoi{10.1086/381073}

\bibitem[{J. {Bosch} {et~al.}(2018){Bosch}, {Armstrong}, {Bickerton}, {Furusawa}, {Ikeda}, {Koike}, {Lupton}, {Mineo}, {Price}, {Takata}, {Tanaka}, {Yasuda}, {AlSayyad}, {Becker}, {Coulton}, {Coupon}, {Garmilla}, {Huang}, {Krughoff}, {Lang}, {Leauthaud}, {Lim}, {Lust}, {MacArthur}, {Mandelbaum}, {Miyatake}, {Miyazaki}, {Murata}, {More}, {Okura}, {Owen}, {Swinbank}, {Strauss}, {Yamada}, \& {Yamanoi}}]{2018PASJ...70S...5B}
{Bosch}, J., {Armstrong}, R., {Bickerton}, S., {et~al.} 2018, \bibinfo{title}{{The Hyper Suprime-Cam software pipeline},} \pasj, 70, S5, \dodoi{10.1093/pasj/psx080}

\bibitem[{J.~S. {Bullock} \& K.~V. {Johnston}(2005){Bullock} \& {Johnston}}]{2005ApJ...635..931B}
{Bullock}, J.~S., \& {Johnston}, K.~V. 2005, \bibinfo{title}{{Tracing Galaxy Formation with Stellar Halos. I. Methods},} \apj, 635, 931, \dodoi{10.1086/497422}

\bibitem[{F. {Castelli} \& R.~L. {Kurucz}(2003){Castelli} \& {Kurucz}}]{2003IAUS..210P.A20C}
{Castelli}, F., \& {Kurucz}, R.~L. 2003, in Modelling of Stellar Atmospheres, ed. N.~{Piskunov}, W.~W. {Weiss}, \& D.~F. {Gray}, Vol. 210, A20, \dodoi{10.48550/arXiv.astro-ph/0405087}

\bibitem[{K.~C. {Chambers} {et~al.}(2016){Chambers}, {Magnier}, {Metcalfe}, {Flewelling}, {Huber}, {Waters}, {Denneau}, {Draper}, {Farrow}, {Finkbeiner}, {Holmberg}, {Koppenhoefer}, {Price}, {Rest}, {Saglia}, {Schlafly}, {Smartt}, {Sweeney}, {Wainscoat}, {Burgett}, {Chastel}, {Grav}, {Heasley}, {Hodapp}, {Jedicke}, {Kaiser}, {Kudritzki}, {Luppino}, {Lupton}, {Monet}, {Morgan}, {Onaka}, {Shiao}, {Stubbs}, {Tonry}, {White}, {Ba{\~n}ados}, {Bell}, {Bender}, {Bernard}, {Boegner}, {Boffi}, {Botticella}, {Calamida}, {Casertano}, {Chen}, {Chen}, {Cole}, {Deacon}, {Frenk}, {Fitzsimmons}, {Gezari}, {Gibbs}, {Goessl}, {Goggia}, {Gourgue}, {Goldman}, {Grant}, {Grebel}, {Hambly}, {Hasinger}, {Heavens}, {Heckman}, {Henderson}, {Henning}, {Holman}, {Hopp}, {Ip}, {Isani}, {Jackson}, {Keyes}, {Koekemoer}, {Kotak}, {Le}, {Liska}, {Long}, {Lucey}, {Liu}, {Martin}, {Masci}, {McLean}, {Mindel}, {Misra}, {Morganson}, {Murphy}, {Obaika}, {Narayan}, {Nieto-Santisteban}, {Norberg}, {Peacock}, {Pier}, {Postman}, {Primak}, {Rae}, {Rai}, {Riess}, {Riffeser}, {Rix}, {R{\"o}ser}, {Russel}, {Rutz}, {Schilbach}, {Schultz}, {Scolnic}, {Strolger}, {Szalay}, {Seitz}, {Small}, {Smith}, {Soderblom}, {Taylor}, {Thomson}, {Taylor}, {Thakar}, {Thiel}, {Thilker}, {Unger}, {Urata}, {Valenti}, {Wagner}, {Walder}, {Walter}, {Watters}, {Werner}, {Wood-Vasey}, \& {Wyse}}]{2016arXiv161205560C}
{Chambers}, K.~C., {Magnier}, E.~A., {Metcalfe}, N., {et~al.} 2016, \bibinfo{title}{{The Pan-STARRS1 Surveys},} arXiv e-prints, arXiv:1612.05560, \dodoi{10.48550/arXiv.1612.05560}

\bibitem[{M. {Chiba} {et~al.}(2026){Chiba}, {Wyse}, {Kirby}, {Cohen}, {Dobos}, {Gerasimov}, {Ishigaki}, {Hayashi}, {Filion}, {Arnaboldi}, {Bhattacharya}, {Hirai}, {Kobayashi}, {Komiyama}, {Kuzma}, {Ogami}, {Chies-Santos}, {Klock-Miranda}, {Sestito}, {Budav{\'a}ri}, {Cooper}, {Ding}, {Escala}, {Ferreira}, {Gerhard}, {Henderson}, {Hong}, {Horigome}, {Ikeda}, {Ishikawa}, {Kirihara}, {Li}, {Mardini}, {Martin}, {Miyazaki Sakurako Okamoto}, {Pattnaik}, {Sato}, {Suzuki}, {Szalay}, {Wardana}, {Wei}, {Wu}, {Wu}, {Xu}, {Ye}, {Miki}, {Zhang}, {Zhao}, {Zhao}, \& {Zhao}}]{2026arXiv260409875C}
{Chiba}, M., {Wyse}, R. F.~G., {Kirby}, E.~N., {et~al.} 2026, \bibinfo{title}{{Galactic Archaeology with the Subaru `{\={O}}nohi`ula Prime Focus Spectrograph Strategic Program},} arXiv e-prints, arXiv:2604.09875, \dodoi{10.48550/arXiv.2604.09875}

\bibitem[{A. {Chiti} {et~al.}(2021){Chiti}, {Frebel}, {Simon}, {Erkal}, {Chang}, {Necib}, {Ji}, {Jerjen}, {Kim}, \& {Norris}}]{2021NatAs...5..392C}
{Chiti}, A., {Frebel}, A., {Simon}, J.~D., {et~al.} 2021, \bibinfo{title}{{An extended halo around an ancient dwarf galaxy},} Nature Astronomy, 5, 392, \dodoi{10.1038/s41550-020-01285-w}

\bibitem[{A.~P. {Cooper} {et~al.}(2010){Cooper}, {Cole}, {Frenk}, {White}, {Helly}, {Benson}, {De Lucia}, {Helmi}, {Jenkins}, {Navarro}, {Springel}, \& {Wang}}]{2010MNRAS.406..744C}
{Cooper}, A.~P., {Cole}, S., {Frenk}, C.~S., {et~al.} 2010, \bibinfo{title}{{Galactic stellar haloes in the CDM model},} \mnras, 406, 744, \dodoi{10.1111/j.1365-2966.2010.16740.x}

\bibitem[{M. {Dall'Ora} {et~al.}(2006){Dall'Ora}, {Clementini}, {Kinemuchi}, {Ripepi}, {Marconi}, {Di Fabrizio}, {Greco}, {Rodgers}, {Kuehn}, \& {Smith}}]{2006ApJ...653L.109D}
{Dall'Ora}, M., {Clementini}, G., {Kinemuchi}, K., {et~al.} 2006, \bibinfo{title}{{Variable Stars in the Newly Discovered Milky Way Satellite in Bootes},} \apjl, 653, L109, \dodoi{10.1086/510665}

\bibitem[{A. {Deason} {et~al.}(2014){Deason}, {Wetzel}, \& {Garrison-Kimmel}}]{2014ApJ...794..115D}
{Deason}, A., {Wetzel}, A., \& {Garrison-Kimmel}, S. 2014, \bibinfo{title}{{Satellite Dwarf Galaxies in a Hierarchical Universe: The Prevalence of Dwarf-Dwarf Major Mergers},} \apj, 794, 115, \dodoi{10.1088/0004-637X/794/2/115}

\bibitem[{J. {Ding} {et~al.}(2025){Ding}, {Rockosi}, {Li}, {Koposov}, {Riley}, {Wang}, {Cooper}, {Kizhuprakkat}, {Lambert}, {Medina}, {Sandford}, {Aguilar}, {Ahlen}, {Bianchi}, {Brooks}, {Claybaugh}, {de la Macorra}, {Doel}, {Forero-Romero}, {Gazta{\~n}aga}, {Gontcho A Gontcho}, {Gutierrez}, {Guy}, {Ishak}, {Kehoe}, {Kisner}, {Kremin}, {Lahav}, {Landriau}, {Le Guillou}, {Meisner}, {Miquel}, {Moustakas}, {Prada}, {P{\'e}rez-R{\`a}fols}, {Rossi}, {Sanchez}, {Schubnell}, {Silber}, {Sprayberry}, {Tarl{\'e}}, {Weaver}, \& {Zhou}}]{2025ApJ...994..134D}
{Ding}, J., {Rockosi}, C., {Li}, T.~S., {et~al.} 2025, \bibinfo{title}{{The Draco Dwarf Spheroidal Galaxy in the First Year of Dark Energy Spectroscopic Instrument Data},} \apj, 994, 134, \dodoi{10.3847/1538-4357/ae0a37}

\bibitem[{I. {Ebrov{\'a}} \& E.~L. {{\L}okas}(2017){Ebrov{\'a}} \& {{\L}okas}}]{2017ApJ...850..144E}
{Ebrov{\'a}}, I., \& {{\L}okas}, E.~L. 2017, \bibinfo{title}{{Galaxies with Prolate Rotation in Illustris},} \apj, 850, 144, \dodoi{10.3847/1538-4357/aa96ff}

\bibitem[{R. {Errani} {et~al.}(2015){Errani}, {Penarrubia}, \& {Tormen}}]{2015MNRAS.449L..46E}
{Errani}, R., {Penarrubia}, J., \& {Tormen}, G. 2015, \bibinfo{title}{{Constraining the distribution of dark matter in dwarf spheroidal galaxies with stellar tidal streams.},} \mnras, 449, L46, \dodoi{10.1093/mnrasl/slv012}

\bibitem[{M. {Fellhauer} {et~al.}(2008){Fellhauer}, {Wilkinson}, {Evans}, {Belokurov}, {Irwin}, {Gilmore}, {Zucker}, \& {Kleyna}}]{2008MNRAS.385.1095F}
{Fellhauer}, M., {Wilkinson}, M.~I., {Evans}, N.~W., {et~al.} 2008, \bibinfo{title}{{Modelling the dynamical evolution of the Bootes dwarf spheroidal galaxy},} \mnras, 385, 1095, \dodoi{10.1111/j.1365-2966.2008.12921.x}

\bibitem[{S. {Ferrone} {et~al.}(2023){Ferrone}, {Di Matteo}, {Mastrobuono-Battisti}, {Haywood}, {Snaith}, {Montuori}, {Khoperskov}, \& {Valls-Gabaud}}]{2023A&A...673A..44F}
{Ferrone}, S., {Di Matteo}, P., {Mastrobuono-Battisti}, A., {et~al.} 2023, \bibinfo{title}{{The e-TidalGCs project. Modeling the extra-tidal features generated by Galactic globular clusters},} \aap, 673, A44, \dodoi{10.1051/0004-6361/202244141}

\bibitem[{C. {Filion} \& R.~F.~G. {Wyse}(2021){Filion} \& {Wyse}}]{2021ApJ...923..218F}
{Filion}, C., \& {Wyse}, R. F.~G. 2021, \bibinfo{title}{{The Far-away Blues: Exploring the Furthest Extents of the Bo{\"o}tes I Ultra-faint Dwarf Galaxy},} \apj, 923, 218, \dodoi{10.3847/1538-4357/ac2df1}

\bibitem[{E.~L. {Fitzpatrick}(1999){Fitzpatrick}}]{1999PASP..111...63F}
{Fitzpatrick}, E.~L. 1999, \bibinfo{title}{{Correcting for the Effects of Interstellar Extinction},} \pasp, 111, 63, \dodoi{10.1086/316293}

\bibitem[{D. {Foreman-Mackey}(2016){Foreman-Mackey}}]{2016JOSS....1...24F}
{Foreman-Mackey}, D. 2016, \bibinfo{title}{{corner.py: Scatterplot matrices in Python},} Journal of Open Source Software, 1, 24, \dodoi{10.21105/joss.00024}

\bibitem[{D. {Foreman-Mackey} {et~al.}(2013){Foreman-Mackey}, {Hogg}, {Lang}, \& {Goodman}}]{2013PASP..125..306F}
{Foreman-Mackey}, D., {Hogg}, D.~W., {Lang}, D., \& {Goodman}, J. 2013, \bibinfo{title}{{emcee: The MCMC Hammer},} \pasp, 125, 306, \dodoi{10.1086/670067}

\bibitem[{C.~S. {Frenk} {et~al.}(1988){Frenk}, {White}, {Davis}, \& {Efstathiou}}]{1988ApJ...327..507F}
{Frenk}, C.~S., {White}, S. D.~M., {Davis}, M., \& {Efstathiou}, G. 1988, \bibinfo{title}{{The Formation of Dark Halos in a Universe Dominated by Cold Dark Matter},} \apj, 327, 507, \dodoi{10.1086/166213}

\bibitem[{T.~K. {Fritz} {et~al.}(2018){Fritz}, {Battaglia}, {Pawlowski}, {Kallivayalil}, {van der Marel}, {Sohn}, {Brook}, \& {Besla}}]{2018A&A...619A.103F}
{Fritz}, T.~K., {Battaglia}, G., {Pawlowski}, M.~S., {et~al.} 2018, \bibinfo{title}{{Gaia DR2 proper motions of dwarf galaxies within 420 kpc. Orbits, Milky Way mass, tidal influences, planar alignments, and group infall},} \aap, 619, A103, \dodoi{10.1051/0004-6361/201833343}

\bibitem[{H. {Furusawa} {et~al.}(2018){Furusawa}, {Koike}, {Takata}, {Okura}, {Miyatake}, {Lupton}, {Bickerton}, {Price}, {Bosch}, {Yasuda}, {Mineo}, {Yamada}, {Miyazaki}, {Nakata}, {Koshida}, {Komiyama}, {Utsumi}, {Kawanomoto}, {Jeschke}, {Noumaru}, {Schubert}, {Iwata}, {Finet}, {Fujiyoshi}, {Tajitsu}, {Terai}, \& {Lee}}]{2018PASJ...70S...3F}
{Furusawa}, H., {Koike}, M., {Takata}, T., {et~al.} 2018, \bibinfo{title}{{The on-site quality-assurance system for Hyper Suprime-Cam: OSQAH},} \pasj, 70, S3, \dodoi{10.1093/pasj/psx079}

\bibitem[{ {Gaia Collaboration} {et~al.}(2023){Gaia Collaboration}, {Vallenari}, {Brown}, {Prusti}, {de Bruijne}, {Arenou}, {Babusiaux}, {Biermann}, {Creevey}, {Ducourant}, {Evans}, {Eyer}, {Guerra}, {Hutton}, {Jordi}, {Klioner}, {Lammers}, {Lindegren}, {Luri}, {Mignard}, {Panem}, {Pourbaix}, {Randich}, {Sartoretti}, {Soubiran}, {Tanga}, {Walton}, {Bailer-Jones}, {Bastian}, {Drimmel}, {Jansen}, {Katz}, {Lattanzi}, {van Leeuwen}, {Bakker}, {Cacciari}, {Casta{\~n}eda}, {De Angeli}, {Fabricius}, {Fouesneau}, {Fr{\'e}mat}, {Galluccio}, {Guerrier}, {Heiter}, {Masana}, {Messineo}, {Mowlavi}, {Nicolas}, {Nienartowicz}, {Pailler}, {Panuzzo}, {Riclet}, {Roux}, {Seabroke}, {Sordo}, {Th{\'e}venin}, {Gracia-Abril}, {Portell}, {Teyssier}, {Altmann}, {Andrae}, {Audard}, {Bellas-Velidis}, {Benson}, {Berthier}, {Blomme}, {Burgess}, {Busonero}, {Busso}, {C{\'a}novas}, {Carry}, {Cellino}, {Cheek}, {Clementini}, {Damerdji}, {Davidson}, {de Teodoro}, {Nu{\~n}ez Campos}, {Delchambre}, {Dell'Oro}, {Esquej}, {Fern{\'a}ndez-Hern{\'a}ndez}, {Fraile}, {Garabato}, {Garc{\'\i}a-Lario}, {Gosset}, {Haigron}, {Halbwachs}, {Hambly}, {Harrison}, {Hern{\'a}ndez}, {Hestroffer}, {Hodgkin}, {Holl}, {Jan{\ss}en}, {Jevardat de Fombelle}, {Jordan}, {Krone-Martins}, {Lanzafame}, {L{\"o}ffler}, {Marchal}, {Marrese}, {Moitinho}, {Muinonen}, {Osborne}, {Pancino}, {Pauwels}, {Recio-Blanco}, {Reyl{\'e}}, {Riello}, {Rimoldini}, {Roegiers}, {Rybizki}, {Sarro}, {Siopis}, {Smith}, {Sozzetti}, {Utrilla}, {van Leeuwen}, {Abbas}, {{\'A}brah{\'a}m}, {Abreu Aramburu}, {Aerts}, {Aguado}, {Ajaj}, {Aldea-Montero}, {Altavilla}, {{\'A}lvarez}, {Alves}, {Anders}, {Anderson}, {Anglada Varela}, {Antoja}, {Baines}, {Baker}, {Balaguer-N{\'u}{\~n}ez}, {Balbinot}, {Balog}, {Barache}, {Barbato}, {Barros}, {Barstow}, {Bartolom{\'e}}, {Bassilana}, {Bauchet}, {Becciani}, {Bellazzini}, {Berihuete}, {Bernet}, {Bertone}, {Bianchi}, {Binnenfeld}, {Blanco-Cuaresma}, {Blazere}, {Boch}, {Bombrun}, {Bossini}, {Bouquillon}, {Bragaglia}, {Bramante}, {Breedt}, {Bressan}, {Brouillet}, {Brugaletta}, {Bucciarelli}, {Burlacu}, {Butkevich}, {Buzzi}, {Caffau}, {Cancelliere}, {Cantat-Gaudin}, {Carballo}, {Carlucci}, {Carnerero}, {Carrasco}, {Casamiquela}, {Castellani}, {Castro-Ginard}, {Chaoul}, {Charlot}, {Chemin}, {Chiaramida}, {Chiavassa}, {Chornay}, {Comoretto}, {Contursi}, {Cooper}, {Cornez}, {Cowell}, {Crifo}, {Cropper}, {Crosta}, {Crowley}, {Dafonte}, {Dapergolas}, {David}, {David}, {de Laverny}, {De Luise}, \& {De March}}]{2023A&A...674A...1G}
{Gaia Collaboration}, {Vallenari}, A., {Brown}, A.~G.~A., {et~al.} 2023, \bibinfo{title}{{Gaia Data Release 3. Summary of the content and survey properties},} \aap, 674, A1, \dodoi{10.1051/0004-6361/202243940}

\bibitem[{C.~R. {Harris} {et~al.}(2020){Harris}, {Millman}, {van der Walt}, {Gommers}, {Virtanen}, {Cournapeau}, {Wieser}, {Taylor}, {Berg}, {Smith}, {Kern}, {Picus}, {Hoyer}, {van Kerkwijk}, {Brett}, {Haldane}, {Fern{\'a}ndez del R{\'\i}o}, {Wiebe}, {Peterson}, {G{\'e}rard-Marchant}, {Sheppard}, {Reddy}, {Weckesser}, {Abbasi}, {Gohlke}, \& {Oliphant}}]{2020Natur.585..357H}
{Harris}, C.~R., {Millman}, K.~J., {van der Walt}, S.~J., {et~al.} 2020, \bibinfo{title}{{Array programming with NumPy},} \nat, 585, 357, \dodoi{10.1038/s41586-020-2649-2}

\bibitem[{A. {Helmi} {et~al.}(2018){Helmi}, {Babusiaux}, {Koppelman}, {Massari}, {Veljanoski}, \& {Brown}}]{2018Natur.563...85H}
{Helmi}, A., {Babusiaux}, C., {Koppelman}, H.~H., {et~al.} 2018, \bibinfo{title}{{The merger that led to the formation of the Milky Way's inner stellar halo and thick disk},} \nat, 563, 85, \dodoi{10.1038/s41586-018-0625-x}

\bibitem[{D.~W. {Hogg}(1999){Hogg}}]{1999astro.ph..5116H}
{Hogg}, D.~W. 1999, \bibinfo{title}{{Distance measures in cosmology},} arXiv e-prints, astro, \dodoi{10.48550/arXiv.astro-ph/9905116}

\bibitem[{S. {Huang} {et~al.}(2018){Huang}, {Leauthaud}, {Murata}, {Bosch}, {Price}, {Lupton}, {Mandelbaum}, {Lackner}, {Bickerton}, {Miyazaki}, {Coupon}, \& {Tanaka}}]{2018PASJ...70S...6H}
{Huang}, S., {Leauthaud}, A., {Murata}, R., {et~al.} 2018, \bibinfo{title}{{Characterization and photometric performance of the Hyper Suprime-Cam Software Pipeline},} \pasj, 70, S6, \dodoi{10.1093/pasj/psx126}

\bibitem[{J.~D. {Hunter}(2007){Hunter}}]{2007CSE.....9...90H}
{Hunter}, J.~D. 2007, \bibinfo{title}{{Matplotlib: A 2D Graphics Environment},} Computing in Science \& Engineering, 9, 90, \dodoi{10.1109/MCSE.2007.55}

\bibitem[{J. {Jensen} {et~al.}(2024){Jensen}, {Hayes}, {Sestito}, {McConnachie}, {Waller}, {Smith}, {Navarro}, \& {Venn}}]{2024MNRAS.527.4209J}
{Jensen}, J., {Hayes}, C.~R., {Sestito}, F., {et~al.} 2024, \bibinfo{title}{{Small-scale stellar haloes: detecting low surface brightness features in the outskirts of Milky Way dwarf satellites},} \mnras, 527, 4209, \dodoi{10.1093/mnras/stad3322}

\bibitem[{K.~V. {Johnston} {et~al.}(2002){Johnston}, {Choi}, \& {Guhathakurta}}]{2002AJ....124..127J}
{Johnston}, K.~V., {Choi}, P.~I., \& {Guhathakurta}, P. 2002, \bibinfo{title}{{Interpreting the Morphology of Diffuse Light around Satellite Galaxies},} \aj, 124, 127, \dodoi{10.1086/341040}

\bibitem[{S. {Kawanomoto} {et~al.}(2018){Kawanomoto}, {Uraguchi}, {Komiyama}, {Miyazaki}, {Furusawa}, {Finet}, {Hattori}, {Wang}, {Yasuda}, \& {Suzuki}}]{2018PASJ...70...66K}
{Kawanomoto}, S., {Uraguchi}, F., {Komiyama}, Y., {et~al.} 2018, \bibinfo{title}{{Hyper Suprime-Cam: Filters},} \pasj, 70, 66, \dodoi{10.1093/pasj/psy056}

\bibitem[{I. {King}(1962){King}}]{1962AJ.....67..471K}
{King}, I. 1962, \bibinfo{title}{{The structure of star clusters. I. an empirical density law},} \aj, 67, 471, \dodoi{10.1086/108756}

\bibitem[{E.~N. {Kirby} {et~al.}(2011){Kirby}, {Lanfranchi}, {Simon}, {Cohen}, \& {Guhathakurta}}]{2011ApJ...727...78K}
{Kirby}, E.~N., {Lanfranchi}, G.~A., {Simon}, J.~D., {Cohen}, J.~G., \& {Guhathakurta}, P. 2011, \bibinfo{title}{{Multi-element Abundance Measurements from Medium-resolution Spectra. III. Metallicity Distributions of Milky Way Dwarf Satellite Galaxies},} \apj, 727, 78, \dodoi{10.1088/0004-637X/727/2/78}

\bibitem[{J. {Kleyna} {et~al.}(2002){Kleyna}, {Wilkinson}, {Evans}, {Gilmore}, \& {Frayn}}]{2002MNRAS.330..792K}
{Kleyna}, J., {Wilkinson}, M.~I., {Evans}, N.~W., {Gilmore}, G., \& {Frayn}, C. 2002, \bibinfo{title}{{Dark matter in dwarf spheroidals - II. Observations and modelling of Draco},} \mnras, 330, 792, \dodoi{10.1046/j.1365-8711.2002.05155.x}

\bibitem[{Y. {Komiyama} {et~al.}(2018){Komiyama}, {Obuchi}, {Nakaya}, {Kamata}, {Kawanomoto}, {Utsumi}, {Miyazaki}, {Uraguchi}, {Furusawa}, {Morokuma}, {Uchida}, {Miyatake}, {Mineo}, {Fujimori}, {Aihara}, {Karoji}, {Gunn}, \& {Wang}}]{2018PASJ...70S...2K}
{Komiyama}, Y., {Obuchi}, Y., {Nakaya}, H., {et~al.} 2018, \bibinfo{title}{{Hyper Suprime-Cam: Camera dewar design},} \pasj, 70, S2, \dodoi{10.1093/pasj/psx069}

\bibitem[{S.~E. {Koposov} {et~al.}(2011){Koposov}, {Gilmore}, {Walker}, {Belokurov}, {Evans}, {Fellhauer}, {Gieren}, {Geisler}, {Monaco}, {Norris}, {Okamoto}, {Pe{\~n}arrubia}, {Wilkinson}, {Wyse}, \& {Zucker}}]{2011ApJ...736..146K}
{Koposov}, S.~E., {Gilmore}, G., {Walker}, M.~G., {et~al.} 2011, \bibinfo{title}{{Accurate Stellar Kinematics at Faint Magnitudes: Application to the Bo{\"o}tes I Dwarf Spheroidal Galaxy},} \apj, 736, 146, \dodoi{10.1088/0004-637X/736/2/146}

\bibitem[{Z.-Z. {Li} {et~al.}(2020){Li}, {Qian}, {Han}, {Li}, {Wang}, \& {Jing}}]{2020ApJ...894...10L}
{Li}, Z.-Z., {Qian}, Y.-Z., {Han}, J., {et~al.} 2020, \bibinfo{title}{{Constraining the Milky Way Mass Profile with Phase-space Distribution of Satellite Galaxies},} \apj, 894, 10, \dodoi{10.3847/1538-4357/ab84f0}

\bibitem[{N. {Longeard} {et~al.}(2022){Longeard}, {Jablonka}, {Arentsen}, {Thomas}, {Aguado}, {Carlberg}, {Lucchesi}, {Malhan}, {Martin}, {McConnachie}, {Navarro}, {S{\'a}nchez-Janssen}, {Sestito}, {Starkenburg}, \& {Yuan}}]{2022MNRAS.516.2348L}
{Longeard}, N., {Jablonka}, P., {Arentsen}, A., {et~al.} 2022, \bibinfo{title}{{The Pristine dwarf galaxy survey - IV. Probing the outskirts of the dwarf galaxy Bo{\"o}tes I},} \mnras, 516, 2348, \dodoi{10.1093/mnras/stac1827}

\bibitem[{K. {Malhan} {et~al.}(2021){Malhan}, {Yuan}, {Ibata}, {Arentsen}, {Bellazzini}, \& {Martin}}]{2021ApJ...920...51M}
{Malhan}, K., {Yuan}, Z., {Ibata}, R.~A., {et~al.} 2021, \bibinfo{title}{{Evidence of a Dwarf Galaxy Stream Populating the Inner Milky Way Halo},} \apj, 920, 51, \dodoi{10.3847/1538-4357/ac1675}

\bibitem[{N.~F. {Martin} {et~al.}(2008){Martin}, {de Jong}, \& {Rix}}]{2008ApJ...684.1075M}
{Martin}, N.~F., {de Jong}, J. T.~A., \& {Rix}, H.-W. 2008, \bibinfo{title}{{A Comprehensive Maximum Likelihood Analysis of the Structural Properties of Faint Milky Way Satellites},} \apj, 684, 1075, \dodoi{10.1086/590336}

\bibitem[{N.~F. {Martin} {et~al.}(2022){Martin}, {Venn}, {Aguado}, {Starkenburg}, {Gonz{\'a}lez Hern{\'a}ndez}, {Ibata}, {Bonifacio}, {Caffau}, {Sestito}, {Arentsen}, {Allende Prieto}, {Carlberg}, {Fabbro}, {Fouesneau}, {Hill}, {Jablonka}, {Kordopatis}, {Lardo}, {Malhan}, {Mashonkina}, {McConnachie}, {Navarro}, {S{\'a}nchez-Janssen}, {Thomas}, {Yuan}, \& {Mucciarelli}}]{2022Natur.601...45M}
{Martin}, N.~F., {Venn}, K.~A., {Aguado}, D.~S., {et~al.} 2022, \bibinfo{title}{{A stellar stream remnant of a globular cluster below the metallicity floor},} \nat, 601, 45, \dodoi{10.1038/s41586-021-04162-2}

\bibitem[{A.~W. {McConnachie}(2012){McConnachie}}]{2012AJ....144....4M}
{McConnachie}, A.~W. 2012, \bibinfo{title}{{The Observed Properties of Dwarf Galaxies in and around the Local Group},} \aj, 144, 4, \dodoi{10.1088/0004-6256/144/1/4}

\bibitem[{W. {McKinney}(2010){McKinney}}]{mckinney2010data}
{McKinney}, W. 2010, in Proceedings of the 9th Python in Science Conference, 56--61, \dodoi{10.25080/Majora-92bf1922-00a}

\bibitem[{S. {Miyazaki} {et~al.}(2018){Miyazaki}, {Komiyama}, {Kawanomoto}, {Doi}, {Furusawa}, {Hamana}, {Hayashi}, {Ikeda}, {Kamata}, {Karoji}, {Koike}, {Kurakami}, {Miyama}, {Morokuma}, {Nakata}, {Namikawa}, {Nakaya}, {Nariai}, {Obuchi}, {Oishi}, {Okada}, {Okura}, {Tait}, {Takata}, {Tanaka}, {Tanaka}, {Terai}, {Tomono}, {Uraguchi}, {Usuda}, {Utsumi}, {Yamada}, {Yamanoi}, {Aihara}, {Fujimori}, {Mineo}, {Miyatake}, {Oguri}, {Uchida}, {Tanaka}, {Yasuda}, {Takada}, {Murayama}, {Nishizawa}, {Sugiyama}, {Chiba}, {Futamase}, {Wang}, {Chen}, {Ho}, {Liaw}, {Chiu}, {Ho}, {Lai}, {Lee}, {Jeng}, {Iwamura}, {Armstrong}, {Bickerton}, {Bosch}, {Gunn}, {Lupton}, {Loomis}, {Price}, {Smith}, {Strauss}, {Turner}, {Suzuki}, {Miyazaki}, {Muramatsu}, {Yamamoto}, {Endo}, {Ezaki}, {Ito}, {Kawaguchi}, {Sofuku}, {Taniike}, {Akutsu}, {Dojo}, {Kasumi}, {Matsuda}, {Imoto}, {Miwa}, {Suzuki}, {Takeshi}, \& {Yokota}}]{2018PASJ...70S...1M}
{Miyazaki}, S., {Komiyama}, Y., {Kawanomoto}, S., {et~al.} 2018, \bibinfo{title}{{Hyper Suprime-Cam: System design and verification of image quality},} \pasj, 70, S1, \dodoi{10.1093/pasj/psx063}

\bibitem[{R.~R. {Mu{\~n}oz} {et~al.}(2018){Mu{\~n}oz}, {C{\^o}t{\'e}}, {Santana}, {Geha}, {Simon}, {Oyarz{\'u}n}, {Stetson}, \& {Djorgovski}}]{2018ApJ...860...66M}
{Mu{\~n}oz}, R.~R., {C{\^o}t{\'e}}, P., {Santana}, F.~A., {et~al.} 2018, \bibinfo{title}{{A MegaCam Survey of Outer Halo Satellites. III. Photometric and Structural Parameters},} \apj, 860, 66, \dodoi{10.3847/1538-4357/aac16b}

\bibitem[{R.~R. {Mu{\~n}oz} {et~al.}(2010){Mu{\~n}oz}, {Geha}, \& {Willman}}]{2010AJ....140..138M}
{Mu{\~n}oz}, R.~R., {Geha}, M., \& {Willman}, B. 2010, \bibinfo{title}{{Turning the Tides on the Ultra-faint Dwarf Spheroidal Galaxies: Coma Berenices and Ursa Major II},} \aj, 140, 138, \dodoi{10.1088/0004-6256/140/1/138}

\bibitem[{R.~R. {Mu{\~n}oz} {et~al.}(2005){Mu{\~n}oz}, {Frinchaboy}, {Majewski}, {Kuhn}, {Chou}, {Palma}, {Sohn}, {Patterson}, \& {Siegel}}]{2005ApJ...631L.137M}
{Mu{\~n}oz}, R.~R., {Frinchaboy}, P.~M., {Majewski}, S.~R., {et~al.} 2005, \bibinfo{title}{{Exploring Halo Substructure with Giant Stars: The Velocity Dispersion Profiles of the Ursa Minor and Draco Dwarf Spheroidal Galaxies at Large Angular Separations},} \apjl, 631, L137, \dodoi{10.1086/497396}

\bibitem[{F. {Muratore} {et~al.}(2025){Muratore}, {Legnardi}, {Milone}, {Mastrobuono-Battisti}, {Cordoni}, {Gorza}, {Lagioia}, {Bortolan}, {Dondoglio}, {Marino}, \& {Ziliotto}}]{2025arXiv251201547M}
{Muratore}, F., {Legnardi}, M.~V., {Milone}, A.~P., {et~al.} 2025, \bibinfo{title}{{Exploring the ultra-faint dwarf Bootes I using JWST and HST: Metallicity distribution and binaries},} arXiv e-prints, arXiv:2512.01547, \dodoi{10.48550/arXiv.2512.01547}

\bibitem[{J.~F. {Navarro} {et~al.}(1997){Navarro}, {Frenk}, \& {White}}]{1997ApJ...490..493N}
{Navarro}, J.~F., {Frenk}, C.~S., \& {White}, S. D.~M. 1997, \bibinfo{title}{{A Universal Density Profile from Hierarchical Clustering},} \apj, 490, 493, \dodoi{10.1086/304888}

\bibitem[{T. {Nguyen} {et~al.}(2026){Nguyen}, {Necib}, {Li}, {Read}, {Ba{\~n}ares-Hern{\'a}ndez}, {Faucher-Gigu{\`e}re}, {Hayashi}, {McKinnon}, {Pace}, {Sandford}, \& {Yang}}]{2026arXiv260626218N}
{Nguyen}, T., {Necib}, L., {Li}, T.~S., {et~al.} 2026, \bibinfo{title}{{Dark Matter in Draco and Bo{\"o}tes I: Hints of a Core in an Ultra-Faint Dwarf from Simulation-Based Inference},} arXiv e-prints, arXiv:2606.26218, \dodoi{10.48550/arXiv.2606.26218}

\bibitem[{M.~A. {Norris} {et~al.}(2014){Norris}, {Kannappan}, {Forbes}, {Romanowsky}, {Brodie}, {Faifer}, {Huxor}, {Maraston}, {Moffett}, {Penny}, {Pota}, {Smith-Castelli}, {Strader}, {Bradley}, {Eckert}, {Fohring}, {McBride}, {Stark}, \& {Vaduvescu}}]{2014MNRAS.443.1151N}
{Norris}, M.~A., {Kannappan}, S.~J., {Forbes}, D.~A., {et~al.} 2014, \bibinfo{title}{{The AIMSS Project - I. Bridging the star cluster-galaxy divide$^{★}${\textdagger}{\textdaggerdbl}{\textsection}{\textparagraph}},} \mnras, 443, 1151, \dodoi{10.1093/mnras/stu1186}

\bibitem[{M. {Odenkirchen} {et~al.}(2001){Odenkirchen}, {Grebel}, {Harbeck}, {Dehnen}, {Rix}, {Newberg}, {Yanny}, {Holtzman}, {Brinkmann}, {Chen}, {Csabai}, {Hayes}, {Hennessy}, {Hindsley}, {Ivezi{\'c}}, {Kinney}, {Kleinman}, {Long}, {Lupton}, {Neilsen}, {Nitta}, {Snedden}, \& {York}}]{2001AJ....122.2538O}
{Odenkirchen}, M., {Grebel}, E.~K., {Harbeck}, D., {et~al.} 2001, \bibinfo{title}{{New Insights on the Draco Dwarf Spheroidal Galaxy from the Sloan Digital Sky Survey: A Larger Radius and No Tidal Tails},} \aj, 122, 2538, \dodoi{10.1086/323715}

\bibitem[{I. {Ogami} {et~al.}(2025){Ogami}, {Tanaka}, {Komiyama}, {Chiba}, {Guhathakurta}, {Kirby}, {Wyse}, {Filion}, {Gilbert}, {Escala}, {Mori}, {Kirihara}, {Tanaka}, {Ishigaki}, {Hayashi}, {Lee}, {Sharma}, {Kalirai}, \& {Lupton}}]{2025MNRAS.536..530O}
{Ogami}, I., {Tanaka}, M., {Komiyama}, Y., {et~al.} 2025, \bibinfo{title}{{The structure of the stellar halo of the Andromeda galaxy explored with the NB515 for Subaru/HSC - I. New insights on the stellar halo up to 120 kpc},} \mnras, 536, 530, \dodoi{10.1093/mnras/stae2527}

\bibitem[{A.~B. {Pace} {et~al.}(2022){Pace}, {Erkal}, \& {Li}}]{2022ApJ...940..136P}
{Pace}, A.~B., {Erkal}, D., \& {Li}, T.~S. 2022, \bibinfo{title}{{Proper Motions, Orbits, and Tidal Influences of Milky Way Dwarf Spheroidal Galaxies},} \apj, 940, 136, \dodoi{10.3847/1538-4357/ac997b}

\bibitem[{A.~B. {Pace} {et~al.}(2020){Pace}, {Kaplinghat}, {Kirby}, {Simon}, {Tollerud}, {Mu{\~n}oz}, {C{\^o}t{\'e}}, {Djorgovski}, \& {Geha}}]{2020MNRAS.495.3022P}
{Pace}, A.~B., {Kaplinghat}, M., {Kirby}, E., {et~al.} 2020, \bibinfo{title}{{Multiple chemodynamic stellar populations of the Ursa Minor dwarf spheroidal galaxy},} \mnras, 495, 3022, \dodoi{10.1093/mnras/staa1419}

\bibitem[{R. {Pascale} {et~al.}(2026){Pascale}, {Battaglia}, {Arroyo-Polonio}, {Vasiliev}, {Nipoti}, \& {Thomas}}]{2026arXiv260424855P}
{Pascale}, R., {Battaglia}, G., {Arroyo-Polonio}, J.~M., {et~al.} 2026, \bibinfo{title}{{Multi-component, axisymmetric dynamical models of dSphs based on distribution functions: inferences on dark matter and intermediate-mass black holes in Draco and Ursa Minor},} arXiv e-prints, arXiv:2604.24855, \dodoi{10.48550/arXiv.2604.24855}

\bibitem[{J. {Pe{\~n}arrubia} {et~al.}(2008){Pe{\~n}arrubia}, {Navarro}, \& {McConnachie}}]{2008ApJ...673..226P}
{Pe{\~n}arrubia}, J., {Navarro}, J.~F., \& {McConnachie}, A.~W. 2008, \bibinfo{title}{{The Tidal Evolution of Local Group Dwarf Spheroidals},} \apj, 673, 226, \dodoi{10.1086/523686}

\bibitem[{J. {Pe{\~n}arrubia} {et~al.}(2009){Pe{\~n}arrubia}, {Navarro}, {McConnachie}, \& {Martin}}]{2009ApJ...698..222P}
{Pe{\~n}arrubia}, J., {Navarro}, J.~F., {McConnachie}, A.~W., \& {Martin}, N.~F. 2009, \bibinfo{title}{{The Signature of Galactic Tides in Local Group Dwarf Spheroidals},} \apj, 698, 222, \dodoi{10.1088/0004-637X/698/1/222}

\bibitem[{C. {Pianta} {et~al.}(2022){Pianta}, {Capuzzo-Dolcetta}, \& {Carraro}}]{2022ApJ...939....3P}
{Pianta}, C., {Capuzzo-Dolcetta}, R., \& {Carraro}, G. 2022, \bibinfo{title}{{The Impact of Binaries on the Dynamical Mass Estimate of Dwarf Galaxies},} \apj, 939, 3, \dodoi{10.3847/1538-4357/ac9303}

\bibitem[{A. {Pietrinferni} {et~al.}(2021){Pietrinferni}, {Hidalgo}, {Cassisi}, {Salaris}, {Savino}, {Mucciarelli}, {Verma}, {Silva Aguirre}, {Aparicio}, \& {Ferguson}}]{2021ApJ...908..102P}
{Pietrinferni}, A., {Hidalgo}, S., {Cassisi}, S., {et~al.} 2021, \bibinfo{title}{{Updated BaSTI Stellar Evolution Models and Isochrones. II. {\ensuremath{\alpha}}-enhanced Calculations},} \apj, 908, 102, \dodoi{10.3847/1538-4357/abd4d5}

\bibitem[{T.~A. {Roderick} {et~al.}(2016){Roderick}, {Mackey}, {Jerjen}, \& {Da Costa}}]{2016MNRAS.461.3702R}
{Roderick}, T.~A., {Mackey}, A.~D., {Jerjen}, H., \& {Da Costa}, G.~S. 2016, \bibinfo{title}{{Extended stellar substructure surrounding the Bo{\"o}tes I dwarf spheroidal galaxy},} \mnras, 461, 3702, \dodoi{10.1093/mnras/stw1541}

\bibitem[{N.~R. {Sandford} {et~al.}(2026){Sandford}, {Li}, {Koposov}, {Hayashi}, {Pace}, {Erkal}, {Bovy}, {Da Costa}, {Cullinane}, {Ji}, {Kuehn}, {Lewis}, {Zucker}, {Limberg}, {Medina}, {Simon}, {Yang}, \& {(S}}]{2026ApJ...998...47S}
{Sandford}, N.~R., {Li}, T.~S., {Koposov}, S.~E., {et~al.} 2026, \bibinfo{title}{{Chemodynamics of Bo{\"o}tes I with S$^{5}$: Revised Velocity Gradient, Dark Matter Density, and Galactic Chemical Evolution Constraints},} \apj, 998, 47, \dodoi{10.3847/1538-4357/ae2fe5}

\bibitem[{K.~S. {Sato} {et~al.}(2025{\natexlab{a}}){Sato}, {Okamoto}, {Yagi}, {Komiyama}, {Arimoto}, {Wyse}, {Kirby}, {Chiba}, {Ogami}, \& {Tanaka}}]{2025ApJ...993L...7S}
{Sato}, K.~S., {Okamoto}, S., {Yagi}, M., {et~al.} 2025{\natexlab{a}}, \bibinfo{title}{{The Extended Stellar Distribution in the Outskirts of the Ursa Minor Dwarf Spheroidal Galaxy},} \apjl, 993, L7, \dodoi{10.3847/2041-8213/ae0cb3}

\bibitem[{K.~S. {Sato} {et~al.}(2025{\natexlab{b}}){Sato}, {Komiyama}, {Okamoto}, {Yagi}, {Ogami}, {Tanaka}, {Arimoto}, {Chiba}, {Kirby}, {Wyse}, \& {Mori}}]{2025PASJ...77.1259S}
{Sato}, K.~S., {Komiyama}, Y., {Okamoto}, S., {et~al.} 2025{\natexlab{b}}, \bibinfo{title}{{The star formation and chemical evolution histories of the Ursa Minor dwarf spheroidal galaxy},} \pasj, 77, 1259, \dodoi{10.1093/pasj/psaf106}

\bibitem[{E.~F. {Schlafly} \& D.~P. {Finkbeiner}(2011){Schlafly} \& {Finkbeiner}}]{2011ApJ...737..103S}
{Schlafly}, E.~F., \& {Finkbeiner}, D.~P. 2011, \bibinfo{title}{{Measuring Reddening with Sloan Digital Sky Survey Stellar Spectra and Recalibrating SFD},} \apj, 737, 103, \dodoi{10.1088/0004-637X/737/2/103}

\bibitem[{L. {Searle} \& R. {Zinn}(1978){Searle} \& {Zinn}}]{1978ApJ...225..357S}
{Searle}, L., \& {Zinn}, R. 1978, \bibinfo{title}{{Composition of halo clusters and the formation of the galactic halo.},} \apj, 225, 357, \dodoi{10.1086/156499}

\bibitem[{M. {S{\'e}gall} {et~al.}(2007){S{\'e}gall}, {Ibata}, {Irwin}, {Martin}, \& {Chapman}}]{2007MNRAS.375..831S}
{S{\'e}gall}, M., {Ibata}, R.~A., {Irwin}, M.~J., {Martin}, N.~F., \& {Chapman}, S. 2007, \bibinfo{title}{{Draco, a flawless dwarf galaxy*},} \mnras, 375, 831, \dodoi{10.1111/j.1365-2966.2006.11356.x}

\bibitem[{F. {Sestito} {et~al.}(2023){Sestito}, {Zaremba}, {Venn}, {D'Aoust}, {Hayes}, {Jensen}, {Navarro}, {Jablonka}, {Fern{\'a}ndez-Alvar}, {Glover}, {McConnachie}, \& {Chen{\'e}}}]{2023MNRAS.525.2875S}
{Sestito}, F., {Zaremba}, D., {Venn}, K.~A., {et~al.} 2023, \bibinfo{title}{{The extended 'stellar halo' of the Ursa Minor dwarf galaxy},} \mnras, 525, 2875, \dodoi{10.1093/mnras/stad2427}

\bibitem[{R.~S. {Stobie}(1980){Stobie}}]{1980SPIE..264..208S}
{Stobie}, R.~S. 1980, in Society of Photo-Optical Instrumentation Engineers (SPIE) Conference Series, Vol. 264, Conference on Applications of Digital Image Processing to Astronomy, ed. D.~A. {Elliott}, 208--212, \dodoi{10.1117/12.959806}

\bibitem[{J.~L. {Tonry} {et~al.}(2012){Tonry}, {Stubbs}, {Lykke}, {Doherty}, {Shivvers}, {Burgett}, {Chambers}, {Hodapp}, {Kaiser}, {Kudritzki}, {Magnier}, {Morgan}, {Price}, \& {Wainscoat}}]{2012ApJ...750...99T}
{Tonry}, J.~L., {Stubbs}, C.~W., {Lykke}, K.~R., {et~al.} 2012, \bibinfo{title}{{The Pan-STARRS1 Photometric System},} \apj, 750, 99, \dodoi{10.1088/0004-637X/750/2/99}

\bibitem[{E. {Vasiliev} {et~al.}(2021){Vasiliev}, {Belokurov}, \& {Erkal}}]{2021MNRAS.501.2279V}
{Vasiliev}, E., {Belokurov}, V., \& {Erkal}, D. 2021, \bibinfo{title}{{Tango for three: Sagittarius, LMC, and the Milky Way},} \mnras, 501, 2279, \dodoi{10.1093/mnras/staa3673}

\bibitem[{P. {Virtanen} {et~al.}(2020){Virtanen}, {Gommers}, {Oliphant}, {Haberland}, {Reddy}, {Cournapeau}, {Burovski}, {Peterson}, {Weckesser}, {Bright}, {van der Walt}, {Brett}, {Wilson}, {Millman}, {Mayorov}, {Nelson}, {Jones}, {Kern}, {Larson}, {Carey}, {Polat}, {Feng}, {Moore}, {VanderPlas}, {Laxalde}, {Perktold}, {Cimrman}, {Henriksen}, {Quintero}, {Harris}, {Archibald}, {Ribeiro}, {Pedregosa}, {van Mulbregt}, \& {SciPy 1. 0 Contributors}}]{2020NatMe..17..261V}
{Virtanen}, P., {Gommers}, R., {Oliphant}, T.~E., {et~al.} 2020, \bibinfo{title}{{SciPy 1.0: fundamental algorithms for scientific computing in Python},} Nature Medicine, 17, 261, \dodoi{10.1038/s41592-019-0686-2}

\bibitem[{E. {Vitral} {et~al.}(2025){Vitral}, {Pe{\~n}arrubia}, \& {Walker}}]{2025MNRAS.tmp.1898V}
{Vitral}, E., {Pe{\~n}arrubia}, J., \& {Walker}, M.~G. 2025, \bibinfo{title}{{Signatures of dark subhalos in dwarf spheroidal galaxies: I. Fluctuations in surface density},} \mnras, \dodoi{10.1093/mnras/staf2013}

\bibitem[{E. {Vitral} {et~al.}(2024){Vitral}, {van der Marel}, {Sohn}, {Libralato}, {del Pino}, {Watkins}, {Bellini}, {Walker}, {Besla}, {Pawlowski}, \& {Mamon}}]{2024ApJ...970....1V}
{Vitral}, E., {van der Marel}, R.~P., {Sohn}, S.~T., {et~al.} 2024, \bibinfo{title}{{HSTPROMO Internal Proper-motion Kinematics of Dwarf Spheroidal Galaxies. I. Velocity Anisotropy and Dark Matter Cusp Slope of Draco},} \apj, 970, 1, \dodoi{10.3847/1538-4357/ad571c}

\bibitem[{F. {Waller} {et~al.}(2023){Waller}, {Venn}, {Sestito}, {Jensen}, {Kielty}, {Borukhovetskaya}, {Hayes}, {McConnachie}, \& {Navarro}}]{2023MNRAS.519.1349W}
{Waller}, F., {Venn}, K.~A., {Sestito}, F., {et~al.} 2023, \bibinfo{title}{{The Cosmic Hunt for members in the outskirts of ultra-faint dwarf galaxies: Ursa Major I, Coma Berenices, and Bo{\"o}tes I},} \mnras, 519, 1349, \dodoi{10.1093/mnras/stac3563}

\end{thebibliography}
\bibliographystyle{aasjournalv7.bst}
\appendix
\section{Selection of Bo\"otes~I Member Candidates from {\it Gaia} DR3}
\label{app:gaia_selection}

To investigate whether the low-density features identified in the HSC stellar-density map are associated with Bo\"otes~I, we select candidate member stars from {\it Gaia} DR3 \citep{2023A&A...674A...1G}.
We restrict the sample to sources with $G_{\rm RP}<22$ mag and with finite parallax and proper-motion measurements and uncertainties.

We first remove sources with positive parallaxes by
\begin{equation}
    \varpi-3\sigma_{\varpi}<0,
\end{equation}
where $\varpi$ and $\sigma_{\varpi}$ are the {\it Gaia} DR3 parallax and its uncertainty, respectively.
This criterion removes obvious nearby foreground stars.

We then evaluate the consistency of each source with the systemic proper motion of Bo\"otes~I.
For this selection, we adopt
\begin{equation}
    \boldsymbol{\mu}_{0}
    =
    (\mu_{\alpha*,0},\mu_{\delta,0})
    =
    (-0.554,-1.111)\ \mathrm{mas\,yr^{-1}},
\end{equation}
from \citet{2018A&A...619A.103F} as the systemic proper motion of
Bo\"otes~I.

\begin{equation}
(\boldsymbol{\mu}_{0}-\boldsymbol{\mu}_{i})^{\mathsf T}
\mathbf{C}_{\mu,i}^{-1}
(\boldsymbol{\mu}_{0}-\boldsymbol{\mu}_{i})
< 9,
\end{equation}
For each source \(i\), we require
where \(\boldsymbol{\mu}_{i}\) is the measured proper-motion $(\mu_{\alpha*,i},\mu_{\delta,i})$ and \(\mathbf{C}_{\mu,i}\) is the sum of the covariance matrices of the source and the adopted systemic proper motion. This criterion retains sources consistent with the systemic proper motion within the adopted \(3\sigma\) ellipse.
\begin{figure*}[ht!]
    \begin{center}
        \includegraphics[width=6cm]{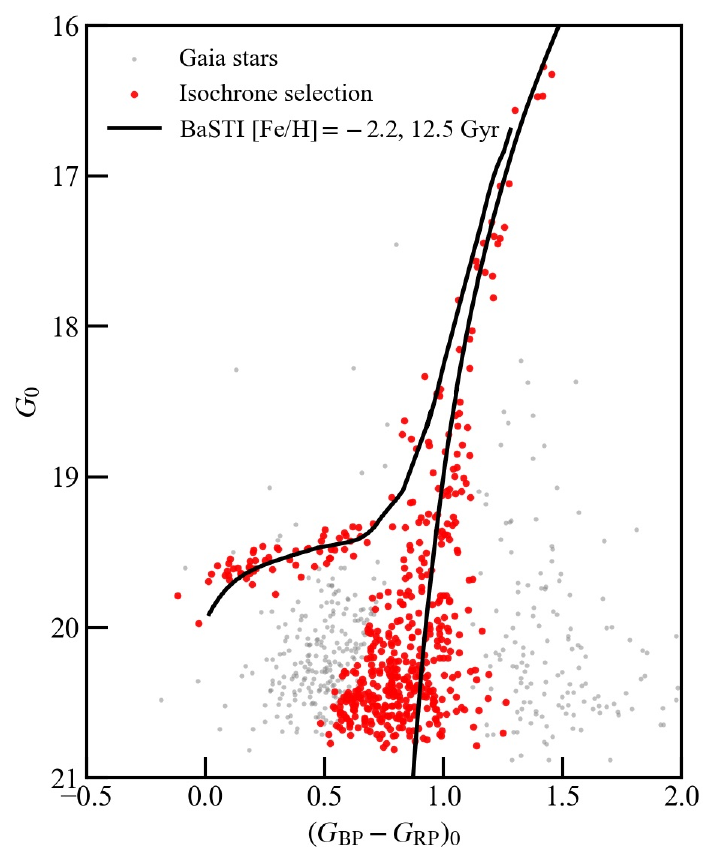}
    \end{center}
    \caption{
        CMD of Bo\"otes~I based on {\it Gaia} DR3.
        The gray points show sources selected using the parallax and proper-motion criteria, including the \(3\sigma\) proper-motion ellipse.
        The red points indicate sources that additionally satisfy the polygonal color--magnitude selection.
        We treat these red sources as likely candidate members of Bo\"otes~I.    }
    \label{fig:Gaia_CMD}
\end{figure*}
Finally, we apply a broad color--magnitude selection as shown in the Figure \ref{fig:Gaia_CMD}.
We further require sufficiently precise {\it Gaia} photometry, retaining only sources with
\begin{equation}
\sigma_G < 0.15~{\rm mag}
\quad {\rm and} \quad
\sigma_{G_{\rm BP}-G_{\rm RP}} < 0.20~{\rm mag}.
\end{equation}
The photometry is corrected for Galactic extinction adopting $E(B-V)=0.0151$ and the extinction coefficients $R_G=2.740$, $R_{G_{\rm BP}}=3.374$, and $R_{G_{\rm RP}}=2.035$.
We compare the extinction-corrected color--magnitude distribution with a BaSTI isochrone with an age of $12.5$ Gyr and ${\rm [Fe/H]}=-2.2$, shifted to a distance of $66$ kpc.
For each source, we calculate the minimum distance to the isochrone in the CMD and evaluate whether this distance is consistent with its photometric-error ellipse.
The uncertainties in $G$ and $G_{\rm BP}-G_{\rm RP}$ are used to weight the magnitude and color differences, respectively. 
To account for the intrinsic metallicity variation of Bo\"otes~I, which broadens the stellar sequence relative to a single isochrone, we add uncertainty of 0.05 mag in quadrature to both the $G$-band and color uncertainties.

\begin{figure*}[ht!]
    \begin{center}
        \includegraphics[width=8cm]{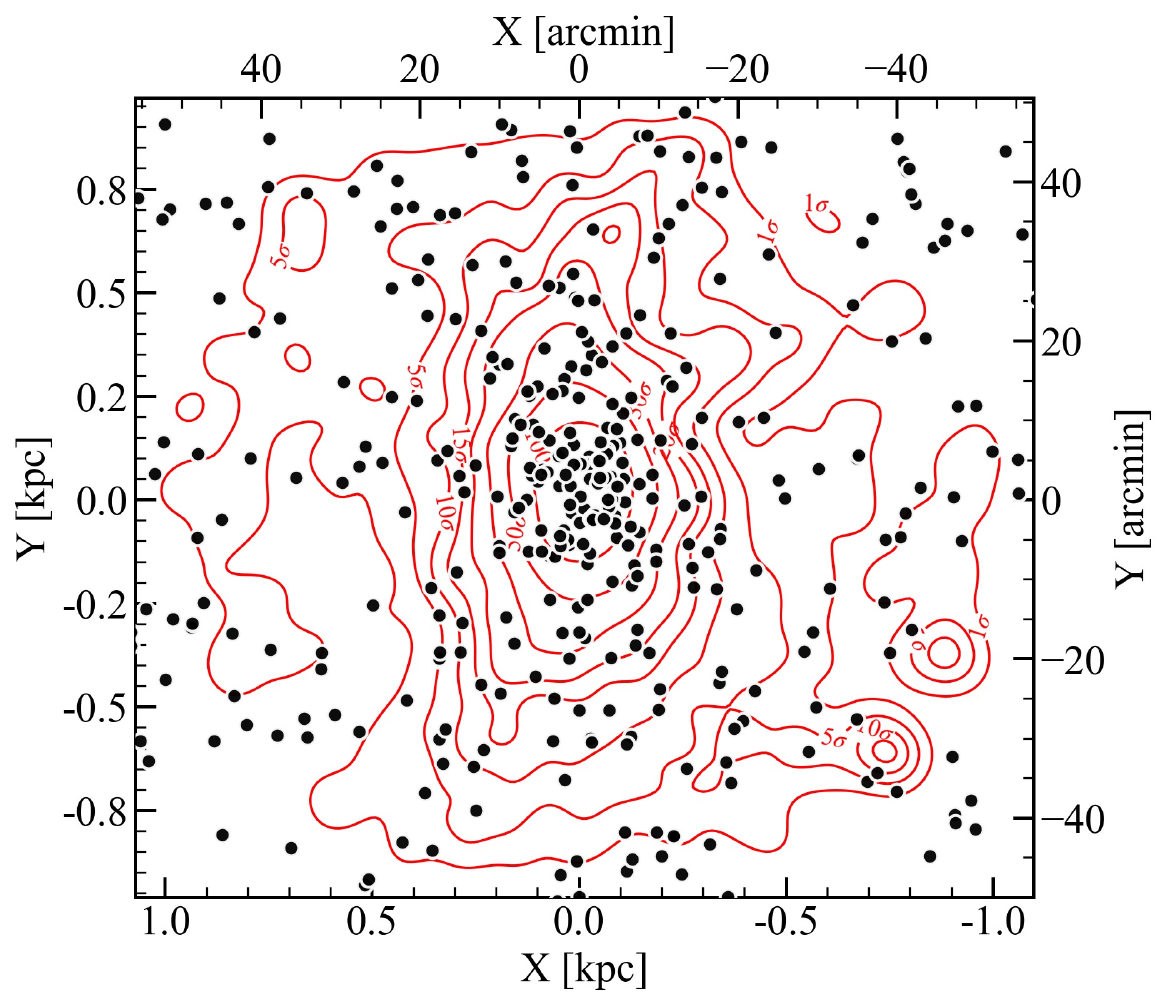}
    \end{center}
    \caption{
        Spatial distribution of the {\it Gaia} DR3 candidate members of Bo\"otes~I.
        The red contours show the stellar-density distribution traced by the HSC-selected MS stars at significance levels of \(1\), \(5\), \(10\), \(15\), \(20\), \(30\), \(50\), and \(100\sigma\) above the background.
        The black points indicate the {\it Gaia} DR3 sources that satisfy the parallax, proper-motion, and color--magnitude selection criteria.
        Several of these candidates are projected along the S-shaped structure and overlap spatially with the southwestern and western clumps identified in the HSC density map.
    }
    \label{fig:Gaia_contour}
\end{figure*}

The resulting candidates therefore satisfy the parallax, proper-motion, and approximate CMD criteria described above.
Their spatial distribution is compared with the stellar-density contours based on the HSC data in Figure~\ref{fig:Gaia_contour}.
The selected candidates show a distribution broadly aligned with the S-shaped structure.
A few of the selected candidates are also seen to overlap spatially with the southwestern clump and the western clump.
Although the small number of selected stars prevents a definitive determination of the origin of these low-density features, their spatial correspondence matches the possibility that they are associated with tidal debris from Bo\"otes~I.
\end{document}